\documentclass[useAMS,usenatbib]{mn2e}
\usepackage{epsfig}
\usepackage{amssymb,subeqnarray}

\begin{document}

\newcommand{\eds}{\Delta\Sigma}
\newcommand{\beq}{\begin{equation}}
\newcommand{\eeq}{\end{equation}}
\newcommand{\bef}{\begin{figure}}
\newcommand{\eef}{\end{figure}}
\newcommand{\hmpc}{$h^{-1}$ Mpc}
\newcommand{\hkpc}{$h^{-1}$ kpc}
\newcommand{\lbcg}{$L_{\rm BCG}$}
\newcommand{\lbcgt}{$L_{{\rm BCG},10}$}
\newcommand{\mlbcg}{$\langle L_{\rm BCG} \rangle$}
\newcommand{\ltot}{$L_{200}$}
\newcommand{\mltot}{$\langle L_{200} \rangle$}   
\newcommand{\ngal}{$N_{200}$} 
\newcommand{\mngal}{$\langle N_{200} \rangle$}
\newcommand{\zbina}{$0.10<z<0.23$}
\newcommand{\zbinb}{$0.23<z<0.30$}
\newcommand{\ds}{$\Delta\Sigma$}
\newcommand{\mnfw}{$M_{200{\bar{\rho}}}$}
\newcommand{\rvir}{$R_{\rm vir}$}
\newcommand{\mpch}{(h^{-1} \mbox{Mpc})^{-3}}
\newcommand{\gpch}{(h^{-1} \mbox{Gpc})^{3}}

\newcommand{\aaps}{{Astron.~Astrophys.~Supp.}}
\newcommand{\araa}{{Annu.~Rev.~Astron.~Astrophys.}}
\newcommand{\aap}{{Astron.~Astrophys.}}
\newcommand{\apjl}{{Astrophys.~J.~Lett.}}
\newcommand{\apj}{{Astrophys.~J.}}
\newcommand{\apjs}{{Astrophys.~J.~Supp.}}
\newcommand{\aj}{{Astron.~J.}}
\newcommand{\prd}{{Phys.~Rev.~D}}
\newcommand{\pasp}{{Pub.~Astron.~Soc.~Pacific}}
\newcommand{\mnras}{{Mon.~Not.~R.~Astron.~Soc.}}
\newcommand{\hMsun}{h^{-1}M_{\odot}}
\newcommand{\physrep}{{Phys. Rep.}}

\title[Improved optical cluster mass tracer]{Improved optical mass tracer for galaxy clusters calibrated using weak lensing measurements}
\author[Reyes et al.]{
  R. Reyes$^1$\thanks{\tt rreyes@astro.princeton.edu}, 
  R. Mandelbaum$^2$\thanks{{\tt rmandelb@ias.edu}, Hubble Fellow}, 
  C. Hirata$^3$, 
  N. Bahcall$^1$, 
  U. Seljak$^{4,5}$
\\$^1$Department of Astrophysical Sciences, Princeton University,
Peyton Hall, Princeton, NJ 08544, USA
\\$^2$Institute for Advanced Study, Einstein Drive, Princeton NJ
  08540, USA 
\\$^3$Mail Code 130-33, Caltech, Pasadena, CA 91125, USA
\\$^4$Institute for Theoretical Physics, University of Zurich, Zurich, Switzerland
\\$^5$Department of Physics, University of California, Berkeley, CA 94720, USA
}


\bibliographystyle{mn2e}
\maketitle

\label{firstpage}
\begin{abstract}

We develop an improved mass tracer for clusters of galaxies from
optically observed parameters, and calibrate the mass relation using
weak gravitational lensing measurements. We employ a sample of $\sim
13~000$ optically-selected clusters from the Sloan Digital Sky Survey
(SDSS) maxBCG catalog, with photometric redshifts in the range
$0.1$--$0.3$. The optical tracers we consider are cluster richness,
cluster luminosity, luminosity of the brightest cluster galaxy (BCG),
and combinations of these parameters. We measure the weak lensing
signal around stacked clusters as a function of the various tracers, and use it to determine the tracer with the least amount of scatter. We further use the weak lensing data to calibrate the mass normalization. We find that the best mass estimator for massive clusters is a combination of cluster richness, \ngal, and the luminosity of the brightest cluster galaxy, \lbcg: $M_{200\bar{\rho}} = (1.27 \pm 0.08) (N_{200}/20)^{1.20 \pm 0.09} (L_{\rm BCG}/\bar{L}_{\rm BCG}(N_{200}))^{0.71 \pm 0.14} \times 10^{14} \hMsun$, where $\bar{L}_{\rm BCG}(N_{200})$ is the observed mean BCG luminosity at a given richness. This improved mass tracer will enable the use of galaxy clusters as a more powerful tool for constraining cosmological parameters.

\end{abstract}


\section{Introduction}
\label{sec:intr}

Clusters of galaxies trace the matter density distribution in
the Universe, and they have long been used successfully as powerful
cosmological probes. Relating the observed cluster abundance to the
dark matter halo abundance predicted by cosmological simulations
provides powerful constraints on a range of cosmological parameters,
including the amplitude of matter fluctuations, neutrino mass and dark
energy density
\citep{1992ApJ...398L..81B,2001ApJ...553..545H,2003NewAR..47..775W,2005PhRvL..95a1302W,2006astro.ph..9591A,2007JCAP...06...24M}.
The strength of these constraints arises from the exponential cutoff
in the cluster mass function for the most massive clusters, which
depends strongly on both the amplitude of matter fluctuations and the matter density.  

Currently, the use of clusters as precise cosmological probes is limited by the lack of reliable mass estimates for a large sample of clusters. While hydrodynamic simulations can provide estimates for the relation between X-ray observable parameters and cluster mass \citep[e.g., ][]{2006ApJ...650..128K,2007ApJ...668....1N}, it is not clear that all the relevant physics determining these relations exist in the simulations. Estimating the virial mass of individual clusters using X-ray measurements \citep[e.g., ][]{2007MNRAS.379..209S} requires the assumption of hydrostatic equilibrium, which introduces potential systematics for non-relaxed clusters, and neglects the effects of non-thermal pressure support, such as that from turbulence, cosmic rays, and magnetic fields. There is a hint of a $\sim 20$ per cent conflict between theoretical predictions and observations for the normalizations of these mass relations \citep{2007A&A...474L..37A,2007ApJ...668....1N}. This discrepancy between hydrostatic masses and total mass also appears to be supported by observational results \citep{2008MNRAS.384.1567M}.Thus, a careful treatment is necessary before they can be used for precision cosmology.

A way to estimate cluster masses that is insensitive to the dynamical state of the system is through weak gravitational lensing measurements. These 
directly probe the total (dark plus luminous) matter distribution. Estimates of the mass of individual clusters using weak lensing are currently 
limited to $\sim 30$ per cent uncertainties by the signal-to-noise ratio of the lensing measurements, for clusters with $M_{500} \sim$ few $\times 10^{14} h^{-1} M_\odot$ \citep[e.g., 
][]{2007MNRAS.379..317H,2007ApJ...667...26P}. They are also subject to systematics such as the shear and source redshift calibration, and limitations due to projection effects of matter near the cluster or along the line-of-sight \citep{2001ApJ...547..560M,2003MNRAS.339.1155H}. These probes can be augmented by strong gravitational lensing measurements \citep{2005A&A...437...49B,2006A&A...458..349C} and velocity dispersion measurements \citep{2007ApJ...669..905B} to aid in the cluster mass determination (e.g., using the methods of \citet{2007ApJ...664..162M} and \citet{2007MNRAS.380.1207S}). 

Here, we calibrate the mass relations for a range of optical
parameters using measurements of the stacked weak lensing signal around a large set of
clusters. This approach is complementary to those methods that provide mass estimates for individual clusters,
which cannot currently be fully applied to large datasets. For example, velocity dispersion measurements 
are limited by the practical difficulty of obtaining spectroscopic observations for a large number of clusters. 
Our method for mass calibration can be readily applied to datasets from upcoming large-scale surveys, such as DES\footnote{https://www.darkenergysurvey.org/}, Pan-STARRS\footnote{http://pan-starrs.ifa.hawaii.edu/public/} and LSST\footnote{http://www.lsst.org/}. 

We employ the largest available sample of $\sim$13~000
galaxy clusters \citep[maxBCG cluster catalog;][]{2007ApJ...660..221K,2007ApJ...660..239K} selected from the Sloan
Digital Sky Survey (SDSS; \citealt{2000AJ....120.1579Y}). 
Stacking the weak lensing signals around many clusters increases the
signal-to-noise ratio that we can achieve. The availability of
accurate photometric redshifts for all objects in the sample
also improve our mass measurements. Independent weak lensing analyses of clusters in this catalog have been performed
\citep{2007arXiv0709.1153S,2007arXiv0709.1159J,2007arXiv0709.1162S}.
Closest to this work is \citet{2007arXiv0709.1159J}, where 
scaling relations of cluster mass with optical richness and cluster luminosity were obtained using a different method for estimating the cluster mass.

In this work, we consider optical tracers available in large cluster surveys, such as cluster richness, cluster luminosity, and luminosity of the brightest cluster galaxy (BCG), and assess how well these parameters trace the cluster mass. In addition, we consider combinations of these parameters and assess whether they provide better mass determinations. Finding the most faithful tracer of cluster mass among the available options will allow us to fully harness the power of clusters in constraining cosmological parameters.

The paper is organized as follows. In Sec.~\ref{sec:data}, we describe the cluster catalog and the weak lensing measurements. In Sec.~\ref{sec:method}, we describe how we use stacked weak lensing measurements to estimate cluster masses, and discuss our approach for assessing mass tracers in Sec.~\ref{sec:interp}. Sec.~\ref{sec:systematics} deals with various tests of systematics. We present our results in Sec.~\ref{sec:results} and conclude in Sec.~\ref{sec:conclusions}.


\begin{figure}
\begin{centering}
\includegraphics[width=75mm]{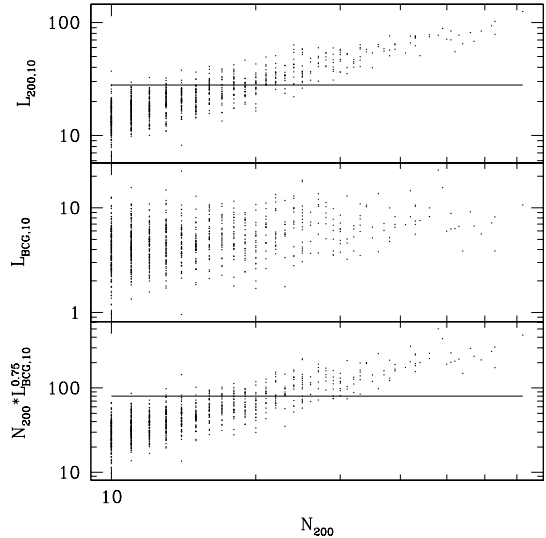}
\caption{Correlation of \ngal {} (cluster richness in red galaxies) with \ltot {} (cluster luminosity in red galaxies), \lbcg {} (BCG luminosity), and the combination \ngal\lbcg$^{0.75}$ (with luminosities in units of $10^{10} h^{-2} L_\odot$). The cluster sample, which is selected by richness ($N_{200}\ge 10$), is complete above $L_{200,10}=30$ and above \ngal$L_{{\rm BCG},10}^{0.75} = 80$ (solid horizontal lines), but is not complete in \lbcg {} at any value.}
\label{fig:correl}
\end{centering}
\end{figure}

\section{Data} \label{sec:data}

In this section, we describe the SDSS data (Sec.~\ref{subsec:data_sdss}), the lens cluster sample from the maxBCG cluster catalog (Sec.~\ref{subsec:data_lens}), and the source galaxy catalog used in the weak lensing analysis (Sec.~\ref{subsec:data_source}).

\subsection{SDSS Data} \label{subsec:data_sdss}

The maxBCG cluster catalog and the lensing source
catalog come from the SDSS, a survey to image roughly
$\pi$ steradians of the sky, and follow up approximately one million
of the detected objects spectroscopically
\citep{2001AJ....122.2267E,2002AJ....123.2945R,2002AJ....124.1810S}.
The imaging is carried out by drift-scanning the sky in photometric
conditions \citep{2001AJ....122.2129H, 2004AN....325..583I}, in five
bands ($ugriz$) \citep{1996AJ....111.1748F,2002AJ....123.2121S} using
a specially-designed wide-field camera
\citep{1998AJ....116.3040G}. These imaging data are used to create the
source catalog that we use in this paper. In addition, objects are
targeted for spectroscopy using these data \citep{2003AJ....125.2276B}
and are observed with a double 320-fiber spectrograph on the same
telescope \citep{2006AJ....131.2332G}. All of these data are processed
by automated pipelines that detect and measure photometric
properties of sources, and astrometrically calibrate the data
\citep{2001adass..10..269L,
  2003AJ....125.1559P,2006AN....327..821T}. The SDSS is nearly complete, 
and has had seven major data releases \citep{2002AJ....123..485S,
  2003AJ....126.2081A, 2004AJ....128..502A, 2005AJ....129.1755A,
  2004AJ....128.2577F,
  2006ApJS..162...38A,2007ApJS..172..634A,2008ApJS..175..297A}.

\subsection{Cluster Lens sample} \label{subsec:data_lens}

Our lens sample consists of 12~612 clusters from the public maxBCG catalog, with richness in red galaxies of $N_{200}\ge 10$ (where the galaxy count includes galaxies brighter than $0.4 L_*$ and located within a scaled radius of $r_{200}$, defined in Eq.~\ref{eq:r200} below). The clusters have photometric redshifts in the range of $z=$ 0.1--0.3, 
selected over a 0.5 $\gpch$ volume covering 7500 deg$^2$ of sky. 
Our sample excludes $\sim 9$ per cent of the solid angle covered by the survey where lensing shape measurements of source galaxies are currently not available.
The maxBCG catalog is presented and discussed in detail by \cite{2007ApJ...660..221K,2007ApJ...660..239K}. 
In this section, we briefly describe the cluster finder algorithm, and define the cluster properties  
used in this work.

The maxBCG cluster finder exploits the existence of the E/S0 red ridgeline of 
cluster galaxies in the color-magnitude diagram, and of a brightest 
cluster galaxy (BCG) found near the centre of most clusters. For each galaxy, it 
obtains a photometric redshift estimate by maximizing the 
likelihood that (i) it is located in an overdensity of E/S0 ridgeline 
galaxies of similar colors, and (ii) it has colors and magnitudes of a 
typical BCG at that redshift. It also determines $N_{1 {\rm Mpc}}$, the 
number of E/S0 ridgeline galaxies located within a projected distance of $1 
h^{-1}$ Mpc of the galaxy, which are dimmer than the galaxy and brighter than $0.4 L^*$, where $L^* = 2.08 \times 10^{10} 
h^{-2} L_\odot$ in the $i$ band at $z=0.1$, with a dependence on redshift 
determined from a Pegase-2 stellar population/galaxy formation model, 
similar to that of \citet{2001AJ....122.2267E}. It then chooses the 
galaxy with the highest likelihood and $N_{1 {\rm Mpc}}$ as a bona fide 
BCG.

To identify cluster members, the cluster size is estimated to be $r_{200}$, the radius within which the galaxy number density of the cluster is 200$\Omega_m^{-1}$ times the mean density of galaxies in the present Universe. The scaled radius $r_{200}$ is estimated from the empirical relation from \cite{2005ApJ...633..122H}:
\beq
r_{200}= 0.156 N_{1 {\rm Mpc}}^{0.6} h^{-1}{\rm Mpc}. \label{eq:r200}
\eeq
The cluster finder identifies galaxies within a scaled radius $r_{200}$ of the BCG, removes them from the list of potential cluster centres, and continues down the list of galaxies with lower likelihood and lower $N_{1 {\rm Mpc}}$ until all candidates are exhausted. For more details, see \citet{2007ApJ...660..221K,2007ApJ...660..239K}.

\cite{2007ApJ...660..221K,2007ApJ...660..239K} performed tests of purity 
and completeness of the maxBCG catalog using mock catalogs from $N$-body 
simulations. They found that the sample is more than 90 per cent pure for clusters with \ngal $\ge 10$; and 90--95 per cent pure for clusters with 
\ngal{}  
$\ge 20$. 
The sample is $>$90 per cent complete for masses $M_{200} \gtrsim 2 \times 10^{14} h^{-1} M_\odot$, and $>95$ per cent complete for masses $M_{200} \gtrsim 
3 
\times 
10^{14} h^{-1} M_\odot$, where $M_{200}$ is the mass within $r_{200}$. These results are of course subject to the assumption that the mock catalogs are a faithful representation of the clusters.

In this work, we use three optical properties of clusters that are reported in the maxBCG catalog:
\begin{itemize}

\item \ngal {} (cluster richness):  the number of E/S0 ridgeline member galaxies fainter than the BCG, brighter than $0.4 L^*$, and located within a projected distance $r_{200}$ (given by Eq.~\ref{eq:r200}) from the BCG. \\

\item \ltot {} (cluster luminosity): the summed $r$ band luminosities of the BCG and the ridgeline member galaxies included in \ngal, $k$-corrected to $z=0.25$. We usually express this luminosity in units of $10^{10} h^{-2} L_\odot$ and denote it by $L_{200,10}$.\\

\item \lbcg {} (BCG luminosity): the $r$ band luminosity of the BCG, $k$-corrected to $z=0.25$. We usually express this luminosity in units of $10^{10} h^{-2} L_\odot$ and denote it by $L_{{\rm BCG},10}$.\\

\end{itemize}
These luminosities are based on SDSS{} `cmodel' magnitudes,
which are constructed from a weighted combination of de Vaucouleurs and
exponential magnitudes. The weights are determined by fitting the galaxy surface brightness profile with a linear combination of the best-fitting de Vaucouleurs and exponential profiles. $K$-corrections are calculated 
from the LRG
template in v4.1.4 of KCORRECT \citep{2003AJ....125.2348B}, using photometric redshifts and without applying a correction for evolution. Galactic extinction correction is
applied using the extinction maps of \cite{1998ApJ...500..525S}. We note that these luminosities may be underestimated (at the 10 per cent level) due 
to systematic errors in sky subtraction, which is most severe in galaxies of large extent \citep{2008ApJS..175..297A}.

Figure~\ref{fig:correl} shows the correlation of the cluster richness
in red galaxies \ngal {} with other optical parameters for the
richness-selected cluster sample ($N_{200}\ge 10$). There is a strong
correlation between \ngal {} and \ltot {} (with a rank correlation
coefficient of 0.68). The sample is complete for cluster luminosities $L_{200,10} \ge 30$ (uppermost panel). On the other hand, 
while the minimum value of \lbcg {} correlates with \ngal, the maximum
value of \lbcg {} does not. The two parameters are weakly correlated,
with rank correlation coefficient is 0.30. The scatter in \lbcg {} at
fixed richness has a Gaussian distribution with width $\gtrsim$ 0.17
dex  \citep{2007arXiv0710.3780H}. The sample is not complete in \lbcg
{} even at the brightest end (middle panel). However, the sample is
complete at \ngal\lbcgt$^{0.75} \ge 80$ (lowermost panel). The
1$\sigma$ statistical error in the luminosities is roughly 0.06 dex (dominated by photometric redshift error), and is much smaller than the observed scatter.

\subsection{Source catalog} \label{subsec:data_source}

The source galaxy sample used for the weak lensing measurements is the same as that originally described in \citet{2005MNRAS.361.1287M}, hereafter M05. This source sample includes over 30 million galaxies from the SDSS imaging data with $r$-band model magnitude brighter than 21.8, with shape measurements obtained using the REGLENS pipeline, including PSF correction done via re-Gaussianization \citep{2003MNRAS.343..459H} and with cuts designed to avoid various shear calibration biases.  A full description of this pipeline can be found in M05.

The REGLENS pipeline obtains galaxy images in the $r$ and $i$ filters from the SDSS ``atlas images'' \citep{2002AJ....123..485S}.  The basic principle of shear measurement using these images is to fit a Gaussian profile with elliptical isophotes to the image, and define the components of the ellipticity 
\beq 
(e_+,e_\times) = \frac{1-(b/a)^2}{1+(b/a)^2}(\cos 2\phi, \sin 2\phi),
\label{eq:e}
\eeq
where $b/a$ is the axis ratio and $\phi$ is the position angle of the major axis.  The ellipticity is then an estimator for the shear, 
\beq
(\gamma_+,\gamma_\times) = \frac{1}{2\cal R}
\langle(e_+,e_\times)\rangle,
\eeq
where ${\cal R}\approx 0.87$ is called the ``shear responsivity'' and 
represents the response of the ellipticity (Eq.~\ref{eq:e}) to a small 
shear \citep{1995ApJ...449..460K, 2002AJ....123..583B}.  In practice, a 
number of corrections need to be applied to obtain the ellipticity.  The 
most important of these is the correction for the smearing and 
circularization of the galactic images by the PSF; M05 uses the PSF maps 
obtained from stellar images by the {\sc psp} pipeline 
\citep{2001adass..10..269L}, and corrects for these using the 
re-Gaussianization technique of \citet{2003MNRAS.343..459H}, which 
includes corrections for non-Gaussianity of both the galaxy profile and 
the PSF.  In order for these corrections to be successful, we require that 
the galaxy be well-resolved compared to the PSF in both $r$ and $i$ bands 
(the only ones used for shape measurement).  To do this we define the 
Gaussian resolution factor:
\beq\label{E:R2def}
R_2 = 1-\frac{T^{(P)}}{T^{(I)}},
\eeq
where the $T$ values are the traces of the adaptive covariance
matrices, and the superscripts indicate whether they are of the PSF or
of the galaxy image.  A large galaxy (compared to the PSF) would have 
$R_2\approx 1$, while a star or other unresolved source would have 
$R_2\approx 0$.  We require that $R_2$ exceed $1/3$
in both $r$ and $i$ bands.

\section{Cluster Masses from Stacked Weak Lensing Measurements} 
\label{sec:method}

In this section, we describe how we estimate cluster masses using
stacked weak lensing measurements. We discuss theory
(Sec.~\ref{sec:theory}), computation of the lensing signal
(Sec.~\ref{sec:signc}), modeling of the density profiles
(Sec.~\ref{sec:density_profiles}), fits to the observed lensing signal
to obtain cluster masses (Sec.~\ref{sec:fits}), and interpretation of
the best-fitting masses (Sec.~\ref{sec:interp}).

\subsection{Theory} \label{sec:theory}

Cluster-galaxy lensing provides a simple way to probe the \
connection between galaxies and matter via their \ cross-correlation function 
\beq
\xi_{gm}(\vec{r}) = \langle \delta_g (\vec{x})
\delta_{m}(\vec{x}+\vec{r})\rangle
\eeq
where $\delta_g$ and $\delta_{m}$ are overdensities of galaxies and matter, respectively.  This cross-correlation can be related to the projected surface density
\beq\label{E:sigmar}
\Sigma(R) = \overline{\rho} \int \left[1+\xi_{gm}\left(\sqrt{R^2 + \chi^2}\right)\right] d\chi
\eeq
(where $r^2=R^2+\chi^2$) which is then related to the observable quantity for lensing, 
\beq \label{E:ds}
\eds(R) = \gamma_t(R) \Sigma_c= \overline{\Sigma}(<R) - \Sigma(R),
\eeq
where $\gamma_t$ is the tangential shear. The second relation is true only 
in the weak lensing limit, for a matter distribution that is axisymmetric 
along the line of sight. This symmetry is naturally achieved by our procedure of 
stacking many clusters and determining their average lensing signal.
This observable quantity can be expressed as the product of the tangential 
shear $\gamma_t$ and a geometric factor
\beq\label{E:sigmacrit}
\Sigma_c = \frac{c^2}{4\pi G} \frac{D_{\rm S}}{D_{\rm L} D_{\rm LS}(1+z_{\rm L})^2},
\eeq
where $D_{\rm L}$ and $D_{\rm S}$ are angular diameter distances to the lens and source, $D_{\rm LS}$ is the angular diameter distance between the lens and source, and the factor of $(1+z_{\rm L})^{-2}$ arises due to our use of comoving coordinates.  For a given lens redshift, $\Sigma_c^{-1}$ rises from zero at $z_{\rm S} = z_{\rm L}$ to an asymptotic value at $z_{\rm S} \gg z_{\rm L}$; that asymptotic value is an increasing function of lens redshift. 

In practice, we truncate the integral in Eq.~\ref{E:sigmar} at the virial radius of the cluster (defined in Eq.~\ref{eq:mvir} below), motivated by attempts to model the lensing signal in simulations (M05). Truncation at two times the virial radius would change the cluster mass estimates at the 5 per cent level.

\subsection{Signal computation} \label{sec:signc}

To compute the average lensing signal \ds$(R)$, lens-source pairs are first assigned weights according to the error on the shape measurement via
\beq
w_{ls} = \frac{\Sigma_c^{-2}}{\sigma_s^2 + \sigma_{SN}^2}
\eeq
where $\sigma_{SN}^2$, the intrinsic shape noise, was determined as a
function of magnitude in M05, Figure 3.  The factor of $\Sigma_c^{-2}$
downweights pairs that are close in redshift, converting the shape
noise in the denominator to a noise in $\Delta\Sigma$.

Once we have computed these weights, we compute the lensing signal in
62 logarithmic radial bins from 0.02 to 9 \hmpc {} as a summation over lens-source pairs via:
\beq
\eds(R) = \frac{\sum_{ls} w_{ls} \gamma_t^{(ls)} \Sigma_c}{2 {\cal 
R}\sum_{ls} w_{ls}},
\eeq
where the factor of 2 arises due to our definition of ellipticity.

There are several additional procedures that must be done when
computing the signal (for more detail, see M05).  First, the signal
computed around random points must be subtracted from the signal
around real lenses to eliminate contributions from systematic
shear. The measured signal around random points is consistent with zero 
over the range of radii we use. Subtraction of this signal 
introduces noise with RMS of $\sim$ 15 per cent on scales from 
0.5 to 1 \hmpc, and $\sim 1$ per cent from 1 to 9 \hmpc. 

Second, the signal must be boosted, i.e., multiplied by $B(R) =
n(R)/n_{\rm rand}(R)$, the ratio of the number density of sources
relative to the number density around random points, in order to account for
the dilution of the lensing signal due to sources that are physically
associated with a lens (i.e., cluster galaxy members), and 
therefore not lensed. 
We find that $B(R)$ decreases with increasing distance from the center, 
ranging from $\sim$ 1.2 to 1.4 at $R=0.5$ \hmpc \ (for low to high-mass 
clusters), and dropping to unity for $R\gtrsim 4$ \hmpc. 

To determine errors on the lensing signal, we divide the
survey area into 200 bootstrap subregions, and generate 2500
bootstrap-resampled datasets. \ Furthermore, to decrease noise in the
covariance matrices due to the bootstrap, we rebin the signal into 22
radial bins (of which 7 are in the range of radii we use for our
fits). 

\subsection{Density profiles} \label{sec:density_profiles}

\begin{figure}
\includegraphics[width=84mm]{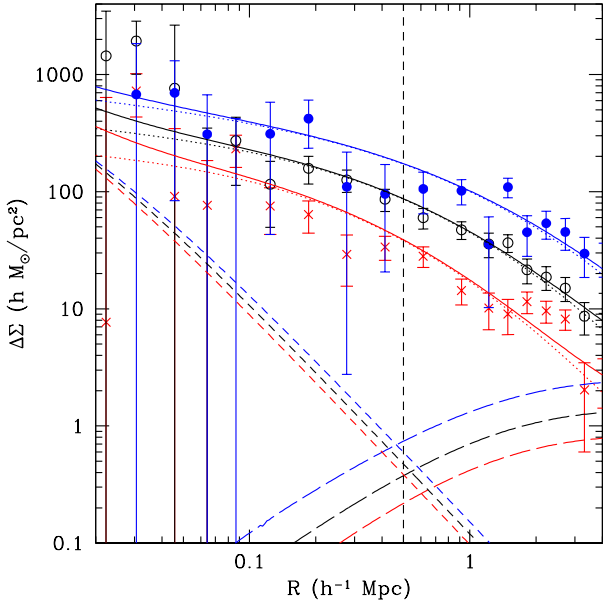}
\caption{Observed mean lensing signals around stacked clusters in three richness bins (data points; from bottom to 
top): \ngal = 10--11, 26--40, and 71--190, with best-fitting masses \mnfw = 
$0.65 \pm 0.30, 2.48 \pm 0.57$, and $8.72 \pm 1.40 \times 10^{14}
h^{-1} M_\odot$, respectively. Also shown are the best-fitting
one-halo and halo-halo profiles (dotted and long-dashed curves,
respectively), the estimated stellar component (short-dashed curves)
and the sum of these three (solid curves). The range of scales used
for the fits is $R=$0.5--4.0 \hmpc {} (rightward of the vertical
dashed line). For this range of scales, the stellar contribution is
negligible and the halo-halo contribution is sub-dominant to the
one-halo term. However, the halo-halo contribution becomes significant
for $R > 1$\hmpc. We model the lensing signal as a sum of the one-halo
and halo-halo profiles.}
\label{fig:ds_n200}
\end{figure}

We model the lensing signal as a sum of contributions from the cluster-mass 
cross-correlation from the cluster (one-halo term) and from large-scale structure 
(halo-halo term). At small scales, contributions from the stars
in the central galaxy are also important, but we show that their contribution is negligible for the range of scales we use for our fits (0.5--4 \hmpc). Figure~\ref{fig:ds_n200} shows the relative contributions of these three components for representative cases. The halo-halo term is significant on scales
$R > 1$\hmpc, but sub-dominant to the one-halo term on all
scales used for the fits.  

The cluster mass distribution is modeled as a Navarro-Frenk-White (hereafter NFW) profile of cold dark matter haloes \citep{1996ApJ...462..563N} 
\beq \label{eq:nfw}
\rho(r)=\frac{\rho_s}{(r/r_s)(1+r/r_s)^2},
\eeq
defined by two parameters, the concentration $c = r_{\rm vir}/r_s$ and the halo 
mass \mnfw. While many definitions are used in the literature, here we
define the virial radius $r_{\rm vir}$ as the radius within which the average density is equal to 200 times the mean density of the Universe $\bar{\rho}$, so that 
\beq \label{eq:mvir}
M_{200{\bar{\rho}}} = \frac{4\pi}{3}r_{\rm vir}^3(200\bar{\rho}),
\eeq 
where the subscript denotes that this mass definition uses $200\bar{\rho}$ rather than the oft-used 200$\rho_{\rm crit}$. The two mass definitions differ by roughly 30 per cent for typical values of concentration. 

We take the concentration to be a fixed function of mass
\beq \label{eq:conc}
c(M_{200{\bar{\rho}}}) = 5.0 \left(\frac{M_{200{\bar{\rho}}}}{10^{14}M_\odot}\right)^{-0.10}.
\eeq
In other words, we assume that the mass distribution only depends on a single parameter, the cluster mass \mnfw. The exponent in Eq.~\ref{eq:conc} matches the results of $N$-body simulations
\citep{2007MNRAS.tmp..922N} and the normalization is determined from the observed density profiles of clusters in the maxBCG catalog \citep{2008arXiv0805.2552M}. 
We find that increasing the normalization from 5.0 to 6.0 results in a
decrease in the best-fitting mass of $\lesssim$ 3 per cent for most of
the mass range we consider. In particular, this means that when we use
a fixed mass-concentration relation, we tend to slightly overestimate
the masses of clusters with high-luminosity BCGs relative to those
that have low-luminosity ones, since the former tend to have earlier
formation times, and therefore, higher concentrations. This effect
would lead to a small positive trend in mass with BCG luminosity at
fixed richness, but we find that the induced slope ($0.025$) is
negligible compared to the observed slopes, $\gamma$ in
Table~\ref{tab:chi2fit}. To estimate this slope, we have used a result
from the simulations of \citep[][; Fig. 4]{2007MNRAS.374.1303C} that indicates that a difference of $\sim$ 1 magnitude in BCG luminosity corresponds to a roughly 20 per cent difference in halo concentration.

The halo-halo contribution to the lensing signal is modeled using the galaxy-matter cross-power 
spectrum as in, e.g., \cite{2005MNRAS.362.1451M}. It is proportional
to the bias $b$, the ratio of the galaxy-matter correlation function
to the matter autocorrelation function. We express the bias as a function of mass or peak height $\nu$ \citep{1999MNRAS.308..119S}:
\beq \label{eq:bias}
b(\nu)=1+\frac{a\nu-1}{\delta_c}+\frac{2p}{\delta_c[1+(a\nu)^p]},
\eeq
where the peak height $\nu = \delta_c^2/\sigma^2(M)$, $\delta_c = 1.686$ 
is the linear overdensity at which a spherical perturbation collapses at 
redshift $z$, and $\sigma(M)$ is the rms fluctuation in spheres that 
contain an average mass $M$ at an initial time, extrapolated using linear 
theory to $z$; we use $z=0.23$, the median redshift of the sample.
For the 
purposes of computing bias, we use $a=0.73$ and $p=0.15$ in order to match 
the results of \cite{2004MNRAS.355..129S}. For example, at $z=0.23$, clusters of mass $6\times 10^{13}$ and $6\times 10^{14} \hMsun$ have biases of 2.2 and 5.5, respectively.

For illustration purposes, we model the stellar component by a Hernquist density profile 
\citep{1990ApJ...356..359H}, which is similar to the NFW profile in 
Eq.~\ref{eq:nfw} but with an exponent of 3 instead of 2, so that it 
falls off faster at large scales. We estimate stellar masses from the mean 
$k$+$e$ corrected $r$ band magnitudes of BCGs in each bin, assuming a 
mass-to-light ratio of $\approx 3 M_\odot/L_\odot$ \citep{2004NewA....9..329P}, following \cite{2006MNRAS.372..758M}. We
estimate the Hernquist profile scale radius by the measured de
Vaucouleurs half-light radius multiplied by a factor of $(\sqrt{2}-1)
\approx 0.414$. Figure~\ref{fig:ds_n200} shows that the stellar contribution to the lensing signal is negligible in the range of scales used for our fits. Thus, we do not include a stellar component in our model of the cluster density profile.

\subsection{Fits to the lensing signal} \label{sec:fits}

We perform fits to the lensing signal at scales $R=0.5$--4.0 \hmpc, which is around the virial radii of clusters in our sample. This choice of fitting range allows us to obtain robust mass estimates (discussed in Sec.~\ref{subsec:sys_offsets}). The stellar contribution to the lensing signal is negligible at these scales (see Fig.~\ref{fig:ds_n200}). We therefore model the lensing signal as a sum of one-halo and halo-halo profiles.

For any \mnfw, we can calculate the one-halo and halo-halo profiles
using
Eqs.~\ref{E:sigmar}, \ref{E:ds}, \ref{eq:nfw}, \ref{eq:conc} and \ref{eq:bias}. Given the observed lensing signal
\ds$(R)$, we determine the best-fitting lensing profile by minimizing
$\chi^2$, using the smooth, analytic (diagonal) covariance matrix. 
We determine formal 1$\sigma$
errors on the best-fitting parameter \mnfw {} using the distribution of parameters obtained from many bootstrap resampled-datasets. This procedure incorporates correlations between the radial bins.

Figure \ref{fig:ds_n200} shows representative examples of observed lensing signals and best-fitting profiles. The halo-halo term becomes important at scales 
$R>1.0$ \hmpc. Neglecting to include this component would yield $\sim$ 7 per cent larger mass estimates compared to fits that include it. 

\subsection{Interpretation of the best-fitting mass} \label{sec:interp}

The stacked weak lensing signal that we measure is the mean signal around a set of
clusters with a range of redshifts and masses. Previous studies
\citep{2005MNRAS.362.1451M}  and the quality of our fits indicate that
the mean signal can be modeled as a single NFW profile to a high degree of accuracy. Moreover, \cite{2005MNRAS.362.1451M} showed that if the mass
distribution is narrow (with a typical width of less than a factor of $\sim 5$ in mass), this model
is able to determine the mean mass of the set of clusters accurately. If there is significant scatter in the mass distribution, then the cluster mass estimate falls between the distribution mean and median.

Here, we consider two kinds of stacking processes: (a) over
a set of clusters that lie within a narrow range of observable properties (e.g., richness or luminosity), and (b) over a set of clusters
that satisfy a threshold in a given property. For case (a), we interpret the
best-fitting mass \mnfw {} as an estimate of the mean mass of the
clusters. We use this approach to calibrate the mean relation between cluster mass and a given cluster observable property.

For case (b), while \mnfw {} may not be a faithful estimate of
the true mean mass because of the broad mass distribution, it nevertheless allows us to assess the relative amount of scatter in a given mass-observable relation
$M= M(O)$. Assuming a monotonic mass-observable relation without scatter,  
rank ordering the clusters by an observable is the same as rank ordering them by mass. 
Thus, selecting the top $N$ clusters by observable would select the $N$ most massive 
clusters. Moreover, if there are two tracers with no scatter they would produce the 
same sample, even if the functional forms $M(O)$ differ. 
The effect of scatter is to bring in clusters with lower mass, which would 
lower the mean weak lensing signal around the stacked clusters and the corresponding best-fitting mass.
Thus, a higher best-fitting mass obtained from a given
observable threshold at fixed number density indicates a lower scatter
in the corresponding mass-observable relation. This analysis has been
worked out explicitly for the case of log normal scatter in
\cite{2007JCAP...06...24M}.  

Finally, we note that the mass that we measure from the weak lensing signal around stacked clusters may differ from other mass definitions, such as from 
spherical overdensity, because the presence of 
substructure and filaments introduce
scatter between the two quantities. This scatter 
may be large if only a small
number of clusters is stacked and one should quantify this with
simulations, which is beyond the scope of this paper. Here we simply take
lensing-defined mass as the mass definition.

\section{Tests of systematics} \label{sec:systematics}

In this section, we discuss various tests of systematics associated with the cluster lens catalog, including photometric 
redshift errors (Sec.~\ref{subsec:sys_photoz}) and offsets from the cluster 
centre (Sec.~\ref{subsec:sys_offsets}), and with the weak lensing source
galaxy catalog, including lensing calibration (Sec.~\ref{subsec:sys_calib}) and contamination from
intrinsic alignments (Sec.~\ref{subsec:sys_ia}).

\subsection{Cluster photometric redshift errors} \label{subsec:sys_photoz}
\begin{figure*}
\includegraphics[width=40mm]{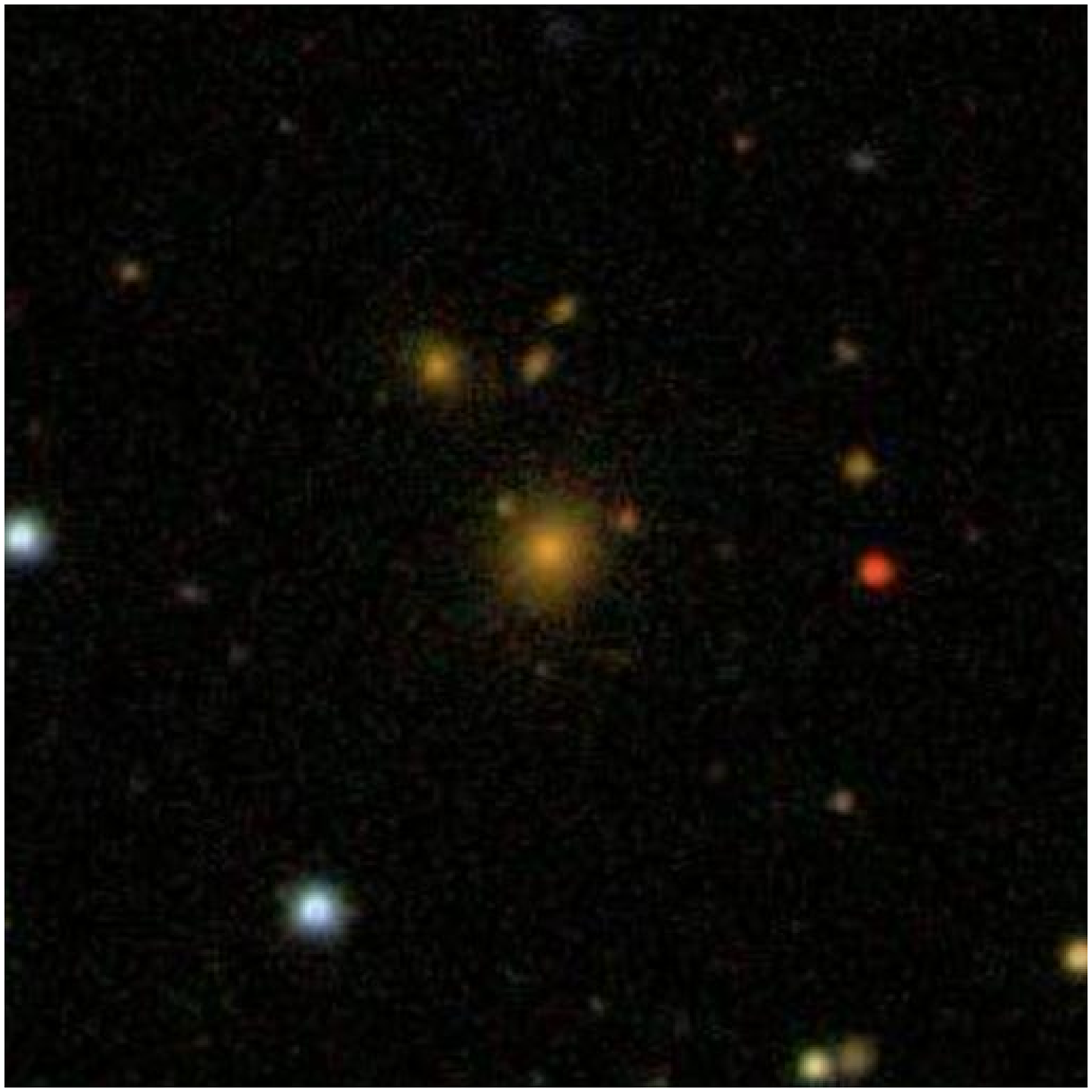}
\includegraphics[width=40mm]{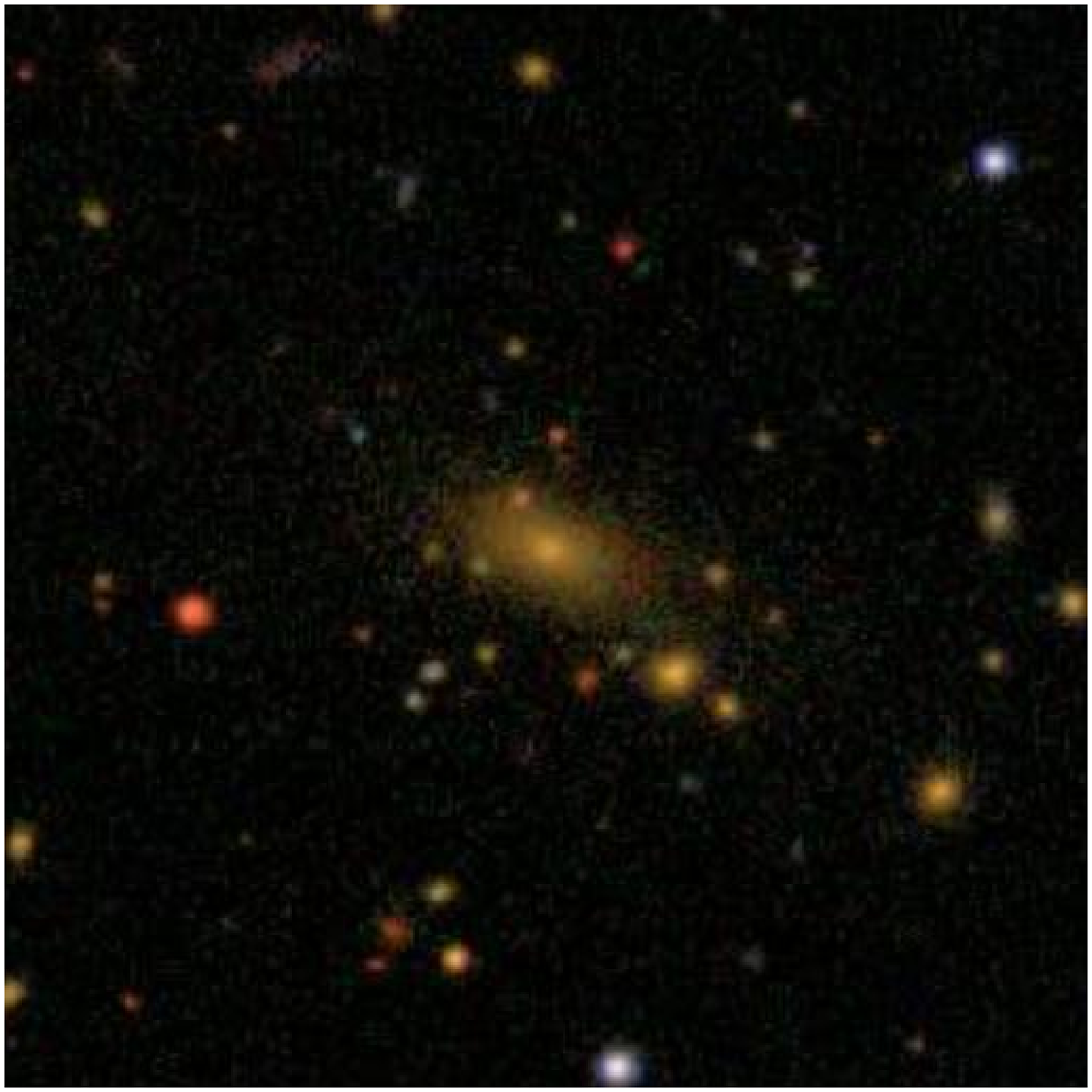}
\includegraphics[width=40mm]{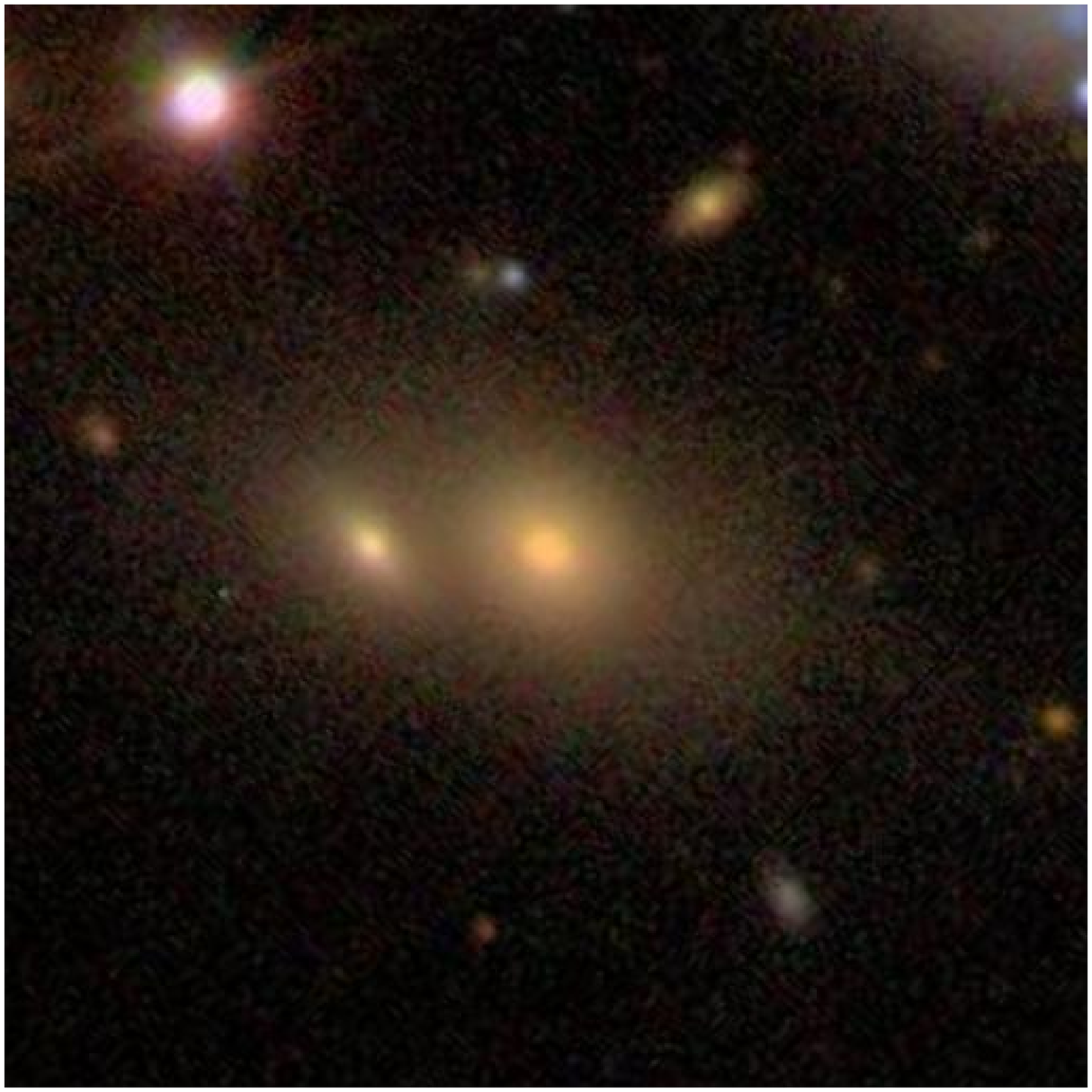}
\includegraphics[width=40mm]{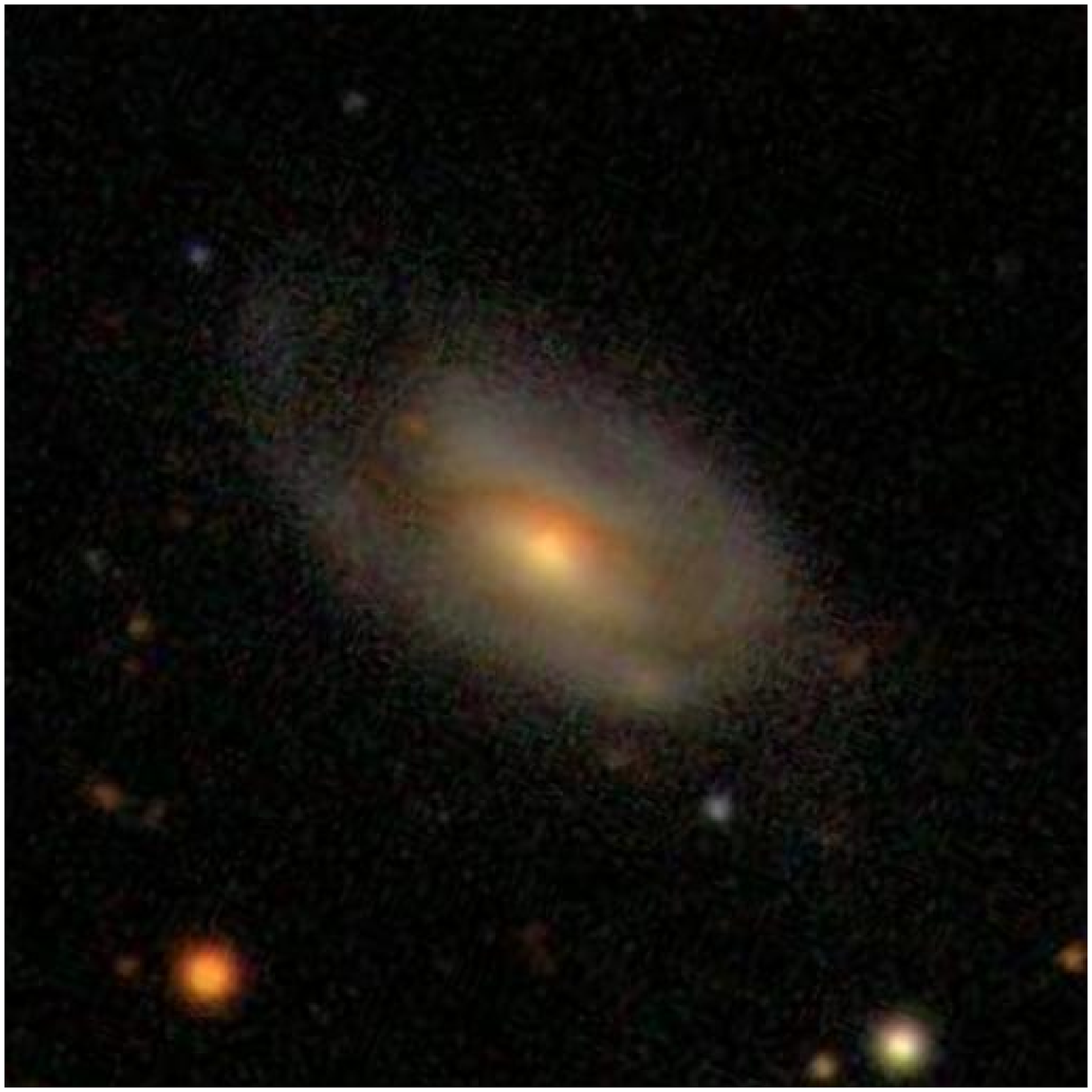}
\caption{Examples of galaxies identified as BCGs in the maxBCG catalog with reported 
$r$ band luminosities ($k$-corrected to $z=0.25$) of \lbcg $>16 \times 10^{10} h^{-2} L_\odot$; these images 
are taken from the SDSS DR6 Skyserver. 
From left to right: (a) SDSS J085540.19-003257.2 ($z=0.271$); 
(b) SDSS J212939.95+000521.1 ($z=0.234$); (c) SDSS J085458.90+490832.3 
($z=0.052$); and (d) SDSS J102246.44+483813.6 ($z=0.050$).  Objects (a) and 
(b) have accurate photo-$z$'s.  These fields show a dominant cD galaxy 
(the BCG) surrounded by other red galaxies, typical of clusters in the 
catalog.  Object (b)'s photo-$z$ was successful despite the presence of
[OII], H$\alpha$, and [NII] emission lines that are unusual for a BCG.
Objects (c) and (d) have severely overestimated photo-$z$'s (0.127 and 
0.138, respectively).  For (c), the error in the photometric 
redshift is probably due to the difficulty in deblending the overlapping 
galaxies. Object (d) seems to be a face-on spiral galaxy with thick 
dust lanes, which was mistaken for a BCG.  We estimate the contamination 
of the catalog from such objects to be $<$2.4 per cent based on the incidence of 
very large errors in photometric redshift for those objects with spectra.}
\label{fig:bcg_panel} 
\end{figure*}

\begin{figure}
\includegraphics[width=84mm]{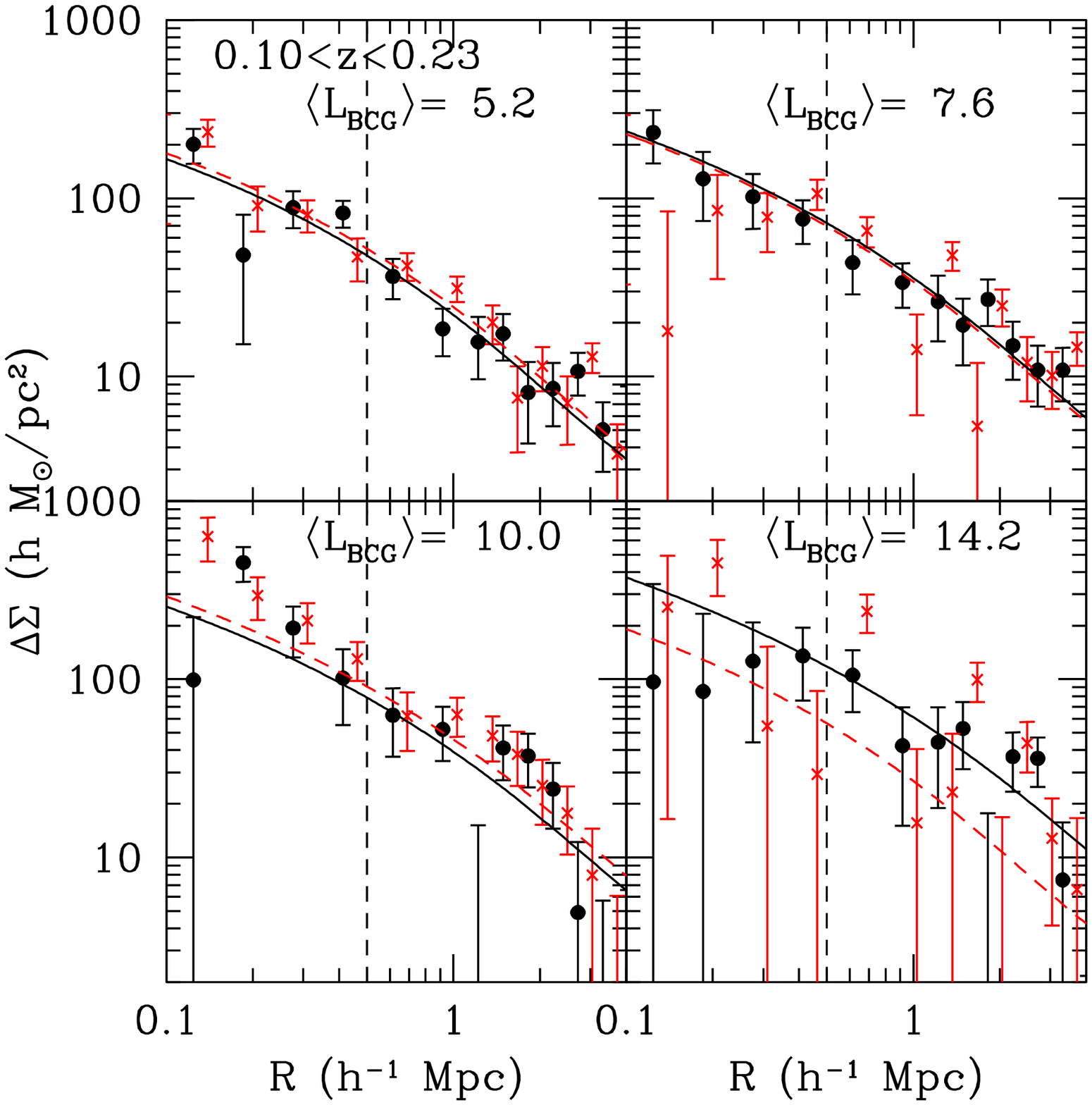}
\includegraphics[width=84mm]{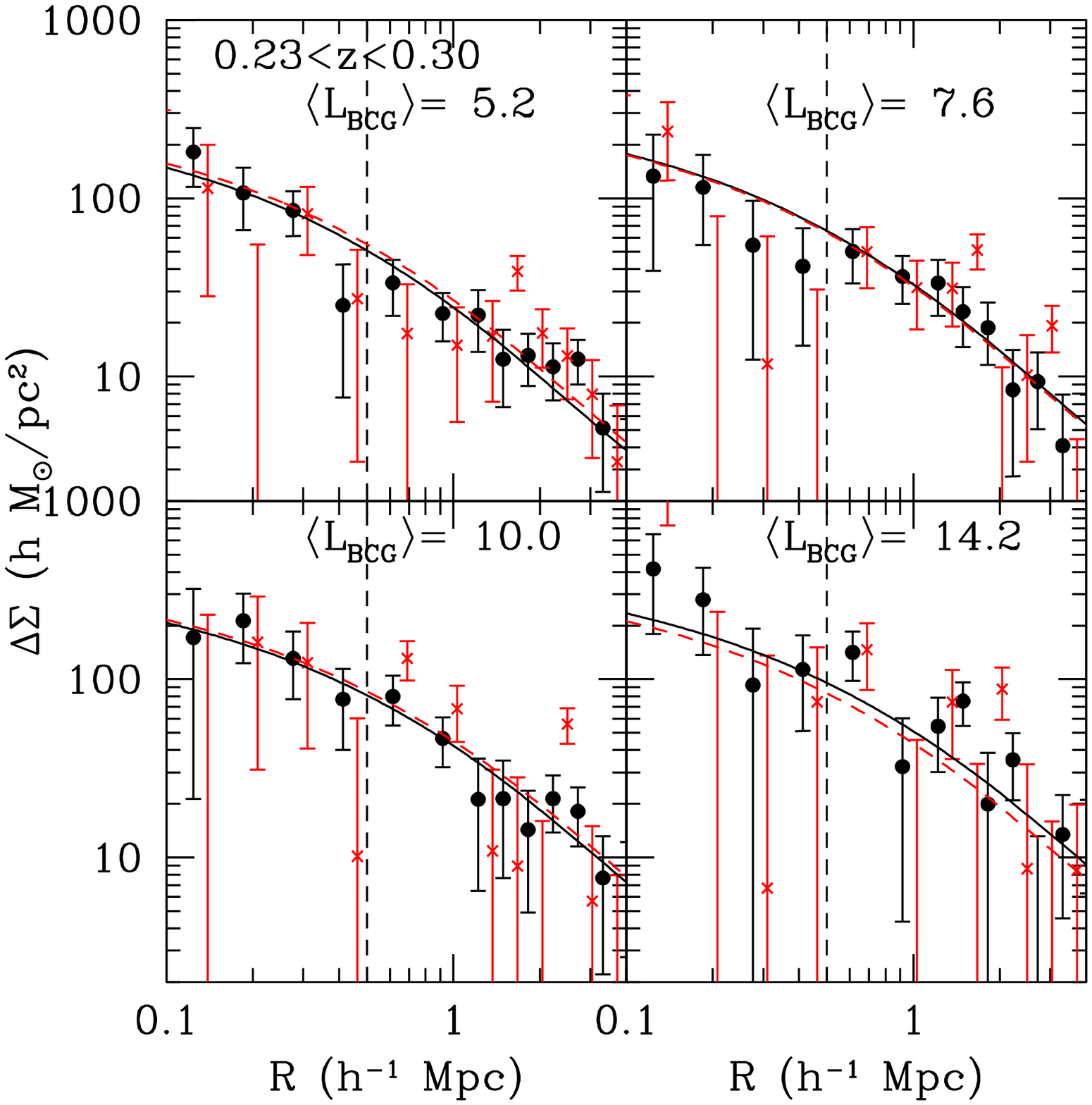}
\caption{Test of systematics for the effect of cluster photo-$z$
errors. Clusters are divided into two ranges in redshift (upper and
lower panels) and five bins in BCG luminosity (the four highest
luminosity bins are shown above, with the mean \lbcg {} listed in
units of $10^{10} h^{-2} L_\odot$). The stacked weak lensing signal
around clusters in each bin is calculated in two ways: (i) using
photometric redshifts and the reported BCG luminosities (filled
circles/black), and (ii) using spectroscopic redshifts and BCG
luminosities scaled to the spectroscopic redshifts (crosses/red). The
best-fitting one-halo + halo-halo profiles are shown in each case
(solid and dashed curves, respectively). The data points 
have been slightly offset horizontally for clarity. The 
vertical dashed line marks the range of scales used in 
our fits $R=$0.5--4.0 \hmpc.}
\label{fig:ds_specz2_z01} 
\end{figure}

\cite{2007ApJ...660..221K,2007ApJ...660..239K} assessed the accuracy of 
photometric redshifts (photo-$z$'s) in the maxBCG catalog by comparing 
them with measured spectroscopic redshifts (available for $\sim$ 40
per cent of the sample). They found that the photo-$z$ dispersion  
$\sqrt{\langle(z_{\rm photo} - z_{\rm spec})^2\rangle}\approx 0.01$, 
and is essentially independent of redshift for the range covered by the sample 
$0.1<z<0.3$. In this section, we investigate the effect of 
photometric redshift errors on our results.

Cluster photo-$z$ errors affect both the measurement of cluster properties
and the computation of the lensing signal. The
reported luminosities in the maxBCG catalog were converted from
apparent magnitudes using distances from photo-$z$'s, so an overestimate in the
redshift would result in a corresponding overestimate in the reported
luminosities. In addition, \ltot {} and \ngal {} would be affected
because the change in both $r_{200}$ and $L_*$ would change which
galaxies would be considered cluster members by the maxBCG cluster finder.  

The lensing signal computation is affected in three ways: first, the 
lensing signal calibration depends on the lens-source geometry, and 
therefore on the assumed value for the cluster redshift; second, the 
conversion from angular distance to transverse separation depends on 
photometric redshift; third, the change in the observed property (luminosity or 
richness) would change the bin in which a given cluster belongs.  
Generically, we expect the first two errors to cancel out at some level 
for any given cluster: e.g., if the lens photo-$z$ is overestimated, then 
$\Sigma_c$ and hence $\Delta\Sigma$ are underestimated, but due to the 
error in the angular diameter distance we also overestimate the transverse 
separation $R$, which increases the signal at fixed transverse 
separation.

Out of 5~423 BCGs (43 per cent of the sample) with measured 
spectroscopic redshifts, 131 galaxies (2.4 per cent) have 
severe photo-$z$ errors, corresponding to differences in distance
moduli larger than 0.5 magnitudes. The incidence of photo-$z$ 
errors is much higher for BCGs with the highest reported luminosities, as 
expected since these extremely luminous objects are rare and a few 
photo-$z$ failures on less luminous objects can lead to a large fractional 
contamination.  Of the 49 objects with reported \lbcg $>16 \times 10^{10} 
h^{-2} L_\odot$, 12 per cent (6 objects) have severe photo-$z$ errors. We show some 
examples in Fig.~\ref{fig:bcg_panel}.

To test for the effect of lens photo-$z$ errors on our weak lensing analysis, we 
divide the 5~423 clusters (with measured spectroscopic redshifts)
into two redshift bins, \zbina {} and \zbinb, 
and five bins in BCG luminosity. We calculate their lensing signal 
in two ways: (i) using photometric redshifts and the reported BCG luminosities, and (ii) using
spectroscopic redshifts and BCG luminosities scaled to the measured spectroscopic redshifts.  Figure~\ref{fig:ds_specz2_z01} 
compares the measured lensing signals for the two cases. Note that 
the binning assignment is different in the two cases because of the 
difference in assumed BCG luminosities. The lensing signals for the 
highest \lbcg {} bins tend to be noisier for case (ii) because these bins 
include very few objects once we correct for photo-$z$ errors. Within the 
error bars, we find no systematic difference between the two cases. Therefore, for our main analysis,
we use the full cluster sample and the reported photometric redshifts 
and luminosities.

\subsection{Offsets from cluster centre}
\label{subsec:sys_offsets}

BCGs are generally expected to lie at or near the centres of clusters, 
where the potential well is the deepest, but 
this is not always observed. Using $N$-body mock galaxy catalogs,
\cite{2007arXiv0709.1159J} found that only $\sim$60--80 per cent of the
BCGs identified by the maxBCG cluster finder are located near the halo
centre, and that the offsets of the rest of the BCGs can be modeled as
a projected Gaussian distribution with a width of 0.42
\hmpc. These results must however be seen in light of the fact that
the halos in the simulations do not correspond exactly to clusters in the data.

For our weak lensing measurements, we define the location of the BCG to be the centre of the cluster, but take steps to reduce the
effect of offsets from the cluster centre on the mass estimates. Fits for the concentration 
from the lensing profiles of clusters in the maxBCG catalog
show that the effect of miscentering is
important (leading to shallower derived concentrations and lower masses) when fits use
transverse separations $R<0.5$ \hmpc, but not when the fits are
restricted to $R>0.5$ \hmpc {} (Mandelbaum, et al 2008). 
Fitting from 0.2 instead of 0.5 \hmpc {} tended to suppress 
the concentrations at the $\sim$ 20 per cent level. Therefore, 
we restrict the fitting range to $R>0.5$ \hmpc {} in this work.

\subsection{Lensing calibration} 
\label{subsec:sys_calib}

Lensing calibration systematics due to the source sample include source redshift uncertainties,
shear calibration, and stellar contamination. Since these effects do
not vary with scale, they could only change the overall normalization in the derived
mass-observable relation.

Comparison with spectroscopy from DEEP2 and zCOSMOS showed that to
account for photometric redshift errors in the source redshifts, 
one has to multiply the signal by a calibration factor of $0.97\pm
0.02$ for the $0.10<z<0.23$ sample, and $0.98\pm 0.04$ for the
$0.23<z<0.30$ sample \citep{2008MNRAS.386..781M}. Stellar
contamination in the source catalog, which would decrease the lensing
signal, is tightly constrained to less than 1 per cent using COSMOS data 
\citep{2008MNRAS.386..781M}. Taking this into account, the
calibration factors become $0.98\pm 0.02$ for the $0.10<z<0.23$ sample
and $0.99\pm 0.04$ for the $0.23<z<0.30$ sample. Since these are
within 1$\sigma$ of unity and are much smaller than the statistical
error bars on the weak lensing signal, we choose not to apply these correction factors
in this work. A conservative estimate of the total calibration
uncertainty, including both these two effects and the shear calibration
bias, is 8 per cent at the 1$\sigma$ level \citep{2005MNRAS.361.1287M}. This 
can be taken into account by adding it in quadrature to the statistical 
error on the mass determinations.

\subsection{Intrinsic alignments}
\label{subsec:sys_ia}

The important intrinsic alignment effect for cluster-galaxy lensing is the 
alignment between the intrinsic ellipticity of a galaxy and the direction 
to nearby cluster BCGs.  This effect comes into play because 
we necessarily include some physically-associated pairs (i.e., pairs of 
lenses and ``sources'' that are really part of the same local structure); 
if these sources preferentially align tangentially or radially 
relative to the lens, they would provide an additive bias to the lensing 
signal.

The effect of intrinsic alignments on the lensing profile is more
important at small transverse separations, since close
physically associated pairs
tend to be more aligned. Using the same source catalog used here, \cite{2006MNRAS.372..758M} 
found that intrinsic alignment
contamination of the lensing signal for luminous red galaxies (LRGs)
is only important at scales $R<0.1$\hmpc, given our procedures for
removing physically associated galaxies from the source sample.  Since
many cluster BCGs are 
also in this LRG sample, this result is relevant for the current work. 
\cite{2006ApJ...644L..25A} measured the mean tangential 
shear of spectroscopically determined satellites and found a tendency
for satellites to align radially towards central galaxies over the
range $7<R<50$ \hkpc. Since we have used photometric redshift estimates
to separate source galaxies from lenses, and we only use the lensing
signal data in the range $R=0.5$--4.0 \hmpc, our results should not be affected by contamination from intrinsic alignments.

Nonetheless, we present constraints on intrinsic alignments contamination of the lensing signal for the
transverse separations used here. To do so, we use the formalism and results on intrinsic alignments
in LRG lenses with our source catalog from
\cite{2006MNRAS.372..758M}. For the ``bright'' lens sample in that work (corresponding to halo masses of
$\sim 7\times 10^{13}$ $h^{-1}M_{\odot}$), the intrinsic alignment
signal was not detected at 0.5--0.6 \hmpc, and was constrained to
contaminate the lensing signal by $<3$ $h M_{\odot} {\rm pc}^{-2}$ at 
95 per cent CL. This constraint is in fact
conservative, since there is reason to believe that the sample of red ``source'' galaxies that we used to place the constraint is more strongly intrinsically aligned than the general galaxy population. Given that the typical lensing signal for the maxBCG clusters on these scales is more than 20 times larger
than this conservative bound, we conclude that it is not an important contaminant for this work. For
larger scales, it is also not important, since the effect is expected to decrease with transverse separation, as does the fraction of physically associated ``source'' galaxies.

\section{Results} 

\label{sec:results}
\begin{figure} 
\includegraphics[width=84mm]{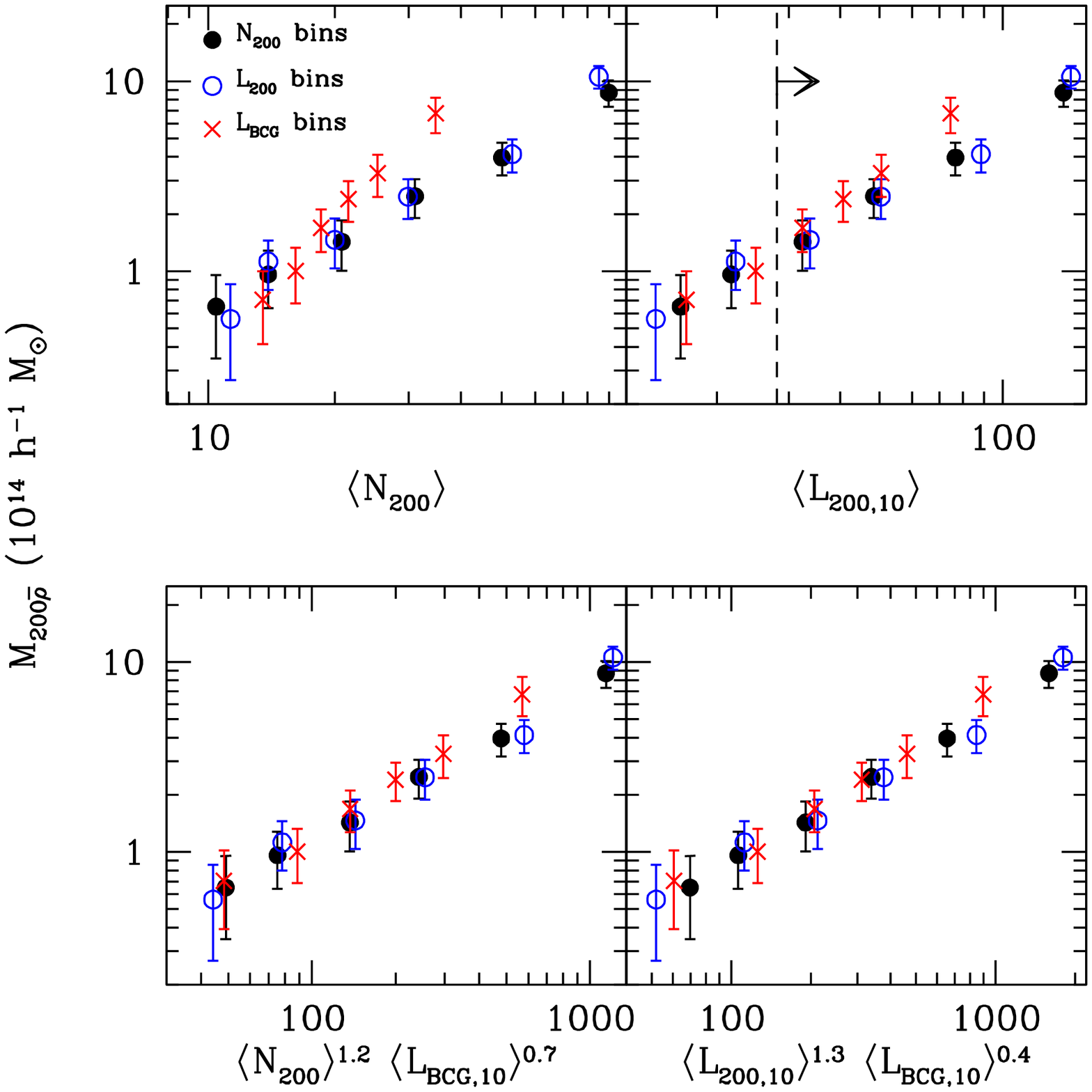}
\caption{Scaling of cluster mass \mnfw {} with various cluster mass tracers. Masses are determined from the stacked weak lensing signal around clusters in individual bins in \ngal {} (filled circles), \ltot {} (open circles), and \lbcg {} (crosses). The upper panels show the scaling of cluster mass with mean parameters $\langle N_{200} \rangle$ and $\langle L_{200} \rangle$; the dashed line on the upper right panel shows the $L_{200}$ value above which the sample is complete. The lower panels show the scaling of cluster mass with a combination of the mean parameters, with exponents taken from Table~\ref{tab:chi2fit} in Sec.~\ref{subsec:results_combined_calib}. The tighter scaling of cluster mass with the combined tracers, regardless of whether \ngal, \ltot, or \lbcg {} is used for the binning, suggests that these combined quantities trace mass more faithfully than either \ngal {} or \ltot {} taken alone.}
\label{fig:mass_bestfit_combined}
\end{figure}

In this section, we calibrate and assess the scatter in the relation between several cluster properties and cluster mass, as outlined in Sec.~\ref{sec:interp}. In Sec.~\ref{subsec:results_indiv}, we consider three main observable parameters-- cluster richness in red galaxies \ngal, cluster luminosity in red galaxies \ltot, and luminosity of the brightest cluster galaxy \lbcg. In Sec.~\ref{subsec:results_combined}, we consider power-law combinations of \ngal {} and \ltot {} with \lbcg, with the aim of finding improved mass tracers for galaxy clusters. 

\subsection{\ngal, \ltot {}  and \lbcg {} as Mass Tracers}
\label{subsec:results_indiv}

\subsubsection{Calibration of mean mass-observable relations}
\label{subsec:results_indiv_calib}
\begin{table*}
\centering
\begin{tabular}{rrrrrr}
\hline
Number  & Range & \mngal & \mltot & \mlbcg & \mnfw \\
\hline
 &  &  & $10^{10} h^{-2} L_\odot$ & $10^{10} h^{-2} L_\odot$ & $10^{14} h^{-1} M_\odot$ \\
\hline
\multicolumn{6}{c}{ Bins in \ngal } \\
\hline
4091 &10--11 &   10.43 & 16.29 &  4.67 & 0.65 $\pm$  0.30 \\ 
5164 &12--17 &   13.88 & 21.67 &  5.27 &   0.96 $\pm$  0.32 \\
2055 &18--25 &   20.78 & 32.38 &  6.21 &   1.43 $\pm$  0.42 \\
 933 &26--40 &   31.06 & 48.40 &  7.05 &   2.48 $\pm$  0.57 \\
 320 &41--70 &   50.06 & 76.64 &  8.24 &   3.96 $\pm$  0.77 \\
  49 &71--190 & 89.86 & 140.87 & 10.45 &   8.72 $\pm$  1.40 \\
\hline
\multicolumn{6}{c}{Bins in \ltot}\\
\hline
4091 & 6.63--17.56  &  11.29 & 14.17 &  3.51 &    0.56 $\pm$  0.29 \\ 
5164 & 17.56--28.51 & 13.89 & 22.22 &  5.57 &     1.12 $\pm$  0.33 \\ 
2055 & 28.51--41.76 & 20.01 & 33.73 &  7.07 &     1.46 $\pm$  0.43 \\ 
 933 & 41.76--64.46 & 29.90 & 50.37 &  8.10 &     2.47 $\pm$  0.59 \\ 
 320 & 64.46--115.55 & 52.95 & 88.44 &  9.83 &    4.13 $\pm$  0.82 \\ 
  49 & 115.55--274.71 & 85.14 & 146.91 & 12.44 & 10.57 $\pm$  1.44 \\ 
\hline
\multicolumn{6}{c}{Bins in \lbcg}\\
\hline
4091 & 0.66--3.95 &  13.47 & 16.84 &  2.95 &   0.71 $\pm$  0.31 \\ 
5164 & 3.95--6.56 &  16.13 & 24.88 &  5.15 &   1.00 $\pm$  0.32 \\ 
2055 & 6.56--8.90 &  18.56 & 32.35 &  7.57 &   1.69 $\pm$  0.42 \\ 
 933 & 8.90--11.73 &  21.56 & 40.73 & 10.02 &  2.40 $\pm$  0.55 \\ 
 320 & 11.74--16.68 & 25.31 & 50.45 & 13.40 &  3.28 $\pm$  0.83 \\ 
  49 & 16.68--29.05 & 34.78 & 74.61 & 19.74 &  6.77 $\pm$  1.57 \\ 
\hline
\end{tabular}
\caption{Individual bins of clusters rank ordered according to: \ngal {} (cluster richness in red galaxies), \ltot {} (cluster luminosity in red galaxies), and \lbcg {} (luminosity of the brightest cluster galaxy). The number of clusters in each bin, their range of properties, mean \ngal, \ltot, \lbcg, and the estimated mean cluster mass \mnfw {} are listed. The 1$\sigma$ errors on the mass estimates are derived from 2500 bootstrap-resampled datasets.}
\label{tab:indiv_allz}
\end{table*}

We begin by calibrating the mean relation between cluster mass and three cluster properties, \ngal {} (cluster richness in red galaxies), \ltot {} (cluster luminosity in red galaxies), and \lbcg {} (luminosity of the brightest cluster galaxy). We rank order the clusters in each property and divide them into six individual bins, keeping the same number of clusters in each bin (Table~\ref{tab:indiv_allz}). We measure the stacked weak lensing signal around clusters in each bin, and determine the best-fitting mass \mnfw {} using the procedure described in Sec.~\ref{sec:fits}. We do this analysis for the full redshift range $0.1<z<0.3$. The results are shown in Table~\ref{tab:indiv_allz} and Fig.~\ref{fig:mass_bestfit_combined}.

The scaling of mean cluster mass with \ngal, \ltot, {} and \lbcg {} are well-described by power laws. To determine the normalization and slope in these relations, we minimize $\chi^2$ simultaneously for the six sets of measured lensing signals. We determine uncertainties on the parameters by repeating the fitting procedure for the 2500 bootstrap-resampled datasets. The best-fitting relations are:
\begin{subeqnarray} \label{eq:scal}
\slabel{eq:scal1}
M_{14}(N_{200})=(1.42\pm 0.08) (N_{200}/20)^{1.16 \pm 0.09} \\
\slabel{eq:scal2}
M_{14}(L_{200})=(1.76\pm 0.17) (L_{200,10}/40)^{1.40 \pm 0.19}  \\
\slabel{eq:scal3} 
M_{14}(L_{\rm BCG})=(1.07\pm 0.07) (L_{{\rm BCG},10}/5)^{1.10 \pm 0.13}  
\end{subeqnarray}
where $M_{14}$ is \mnfw {} in units of $10^{14} h^{-1} M_\odot$, and $L_{200,10}$ and $L_{{\rm BCG},10}$ are in units of $10^{10} h^{-2} L_\odot$. From the 
covariance matrix of the best-fitting parameters, we find that the slope and normalization are uncorrelated for the \ngal {} relation, and anti-correlated at the $\sim 50-60$ per cent level for the \ltot {} and \lbcg {} relations.

The mass-$L_{200}$ relation Eq.~\ref{eq:scal2} is derived using only clusters with $L_{200,10}>28$, where the sample is complete in $L_{200}$ (Fig.~\ref{fig:correl}). Thus, it is not affected by the selection effect introduced by the $N_{200} \ge 10$ cut. If we include the full sample in the analysis, we find $M_{200\bar{\rho},14}(L_{200})=(1.96\pm 0.11)(L_{200,10}/40)^{1.13 \pm 0.08}$. The derived slope is shallower than that in Eq.~\ref{eq:scal2}, consistent with the effect of missing lower mass clusters at low luminosity. The sample is incomplete at all values of \lbcg, so the shallow slope of Eq.~\ref{eq:scal3} is partly due to this selection effect; it should therefore be kept in mind that this relation is valid only for the richness-selected sample. For a sample complete in \lbcg, the slope would likely be steeper and the mean mass at low \lbcg {} would be lower (e.g., \citet{2004ApJ...617..879L} find that BCG luminosity scales with halo mass with an exponent of $0.33\pm 0.06$, which implies a much steeper relation than that in Eq.~\ref{eq:scal3}). 

The slopes we find for the scaling of mass with \ngal {} and \ltot {} are
roughly consistent with the results of \cite{2007arXiv0709.1159J} (Tables
10 and 11 list 1.30 and 1.25, respectively, for the mass definition
closest to ours $M_{180b}$), though it should be noted that they use
additional clusters (with $N_{200} < 10$). Our normalization for the 
mass-richness relation is higher by $\sim$18 per cent, but this can 
be explained by our use of different methods for determining photometric 
redshifts of source galaxies, which leads to different amounts of bias in the estimated 
lensing signals. \citep{2008MNRAS.386..781M} tested for calibration bias in the
lensing signal due to use of different methods of determining source redshifts, using source galaxies with
spectroscopy from zCOSMOS and DEEP2 as a reference. They concluded that for the maxBCG lens redshift
distribution and the methods used here of determining source redshifts, the calibration bias in the
lensing signal is small (consistent with zero within our quoted systematic error), whereas for the SDSS
DR6 neural net photoz's used by \cite{2007arXiv0709.1159J}, it is approximately -18 per cent.

\subsubsection{Scatter in the mass-observable relations}
\label{subsec:results_indiv_scatter}

\begin{figure*}
\includegraphics[width=84mm]{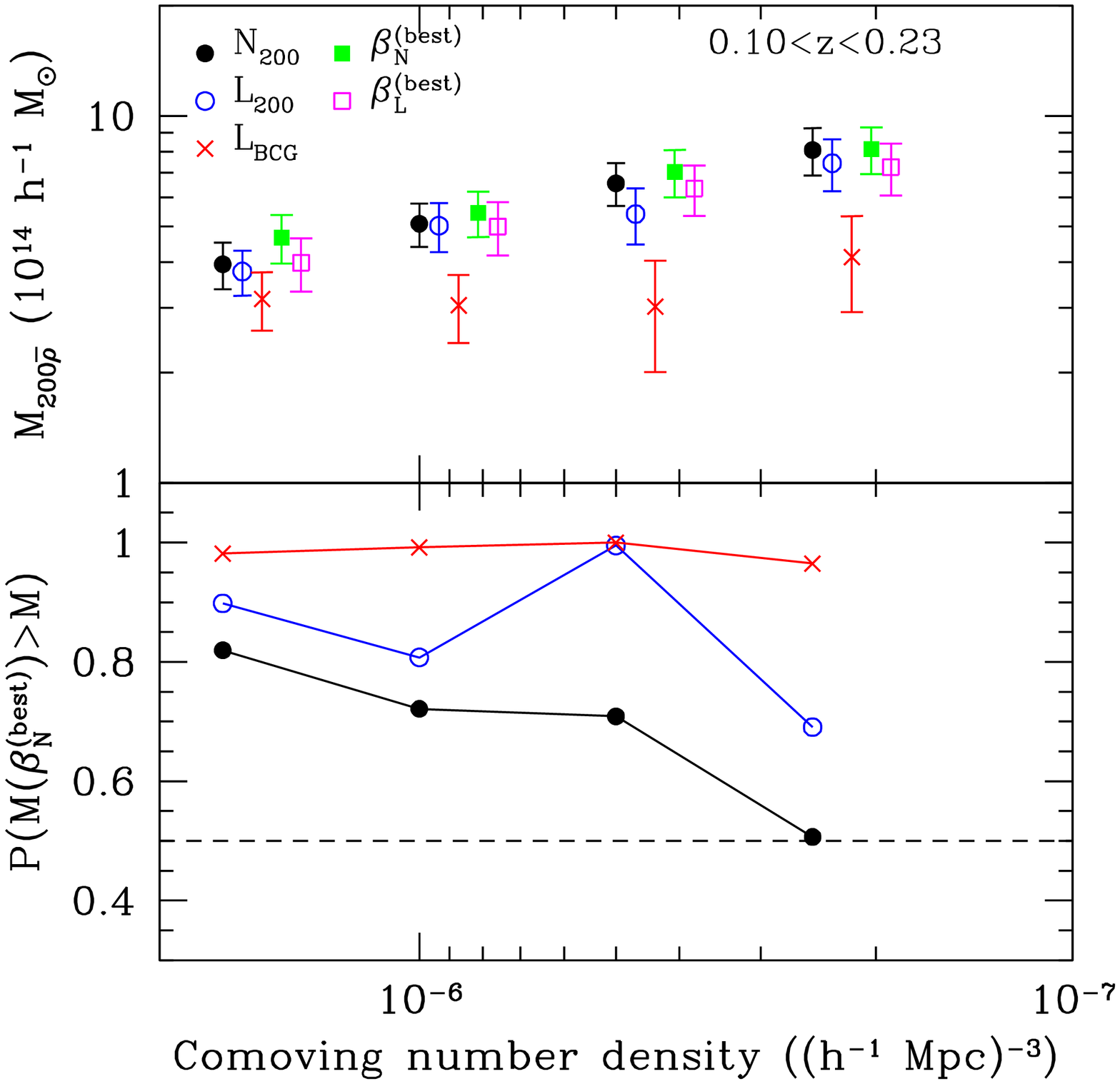}
\includegraphics[width=84mm]{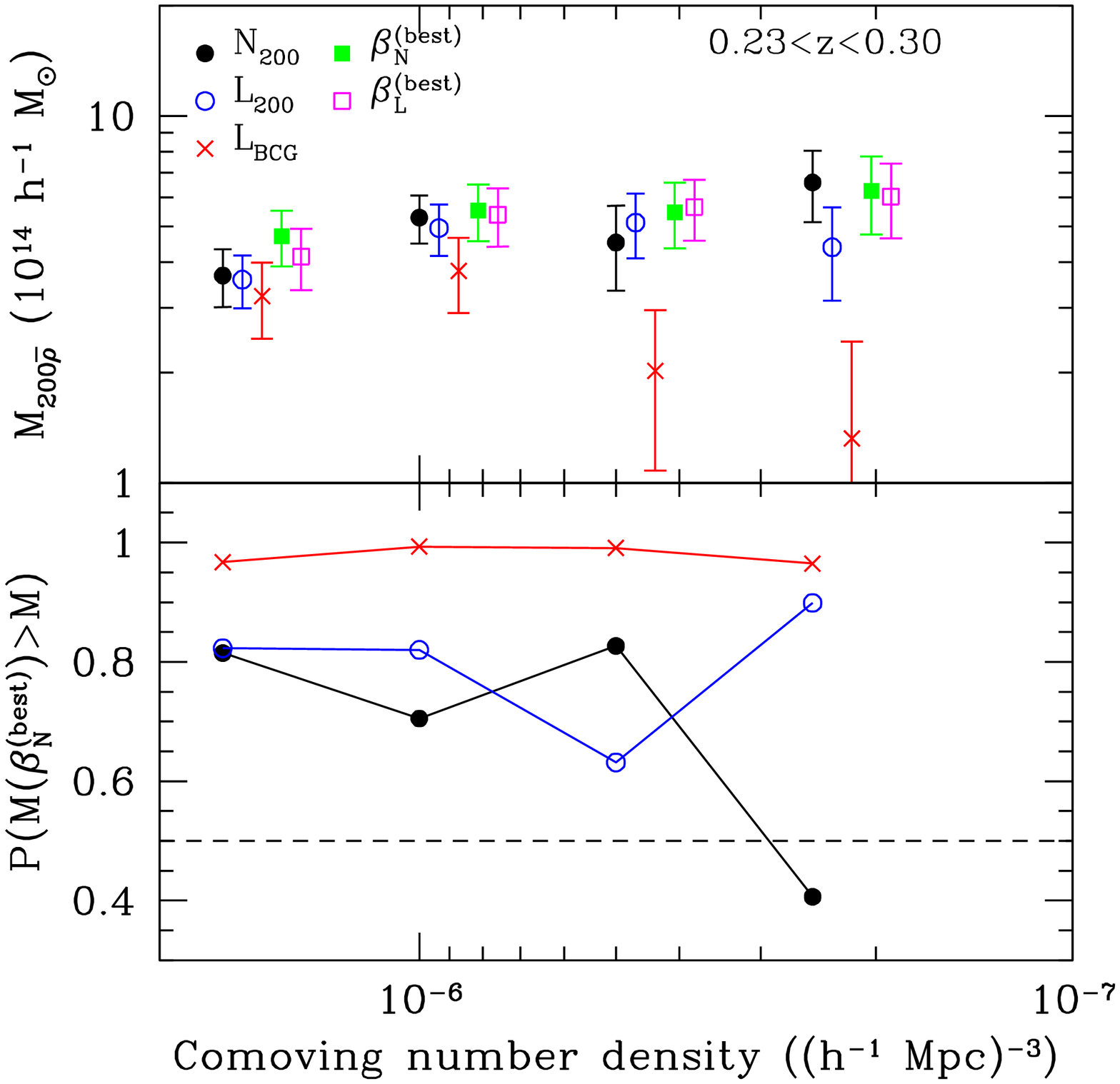}
\caption{Comparison of the relative amount of scatter in the various
mass tracers. Higher values of the best-fitting cluster mass, \mnfw, 
indicate a lower scatter in the mass relation. 
\emph{Upper panels}: Cluster masses \mnfw {} from stacked weak lensing signals around clusters satisfying thresholds in the various tracers,
for comoving number densities $\bar{n} = \{20, 10, 5, \mbox{ and } 2.5\} \times 10^{-7} \mpch$. 
We compare the mass tracers \ngal, \ltot, \lbcg, $N_{200}L_{\rm BCG}^{\beta_N^{\rm (best)}}$ and $L_{200}L_{\rm BCG}^{\beta_L^{\rm (best)}}$
, the combined tracers 
that yield the highest masses at each number density (Sec.~\ref{subsec:results_combined_scatter}). Left and right plots are for the 
two redshift ranges; the data points in the figure are slightly offset horizontally for clarity. 
The 1$\sigma$ error bars shown here are tightly correlated, so the differences in the masses are more significant than apparent by eye. 
\emph{Lower panels}: Probability that the $\beta_N^{\rm (best)}$ tracer yields a higher mass than \ngal {} (filled circles/black), \ltot {} (open circles/blue), or \lbcg {} (crosses/red) taken alone, defined to be the percentage of cases among 1000 bootstrap-resampled datasets. High values of this quantity suggest that the combined tracers have comparable or lower scatter than either \ngal {} or \ltot {} taken alone, for this range of cluster abundances.}
\label{fig:mass_scatter}
\end{figure*}

In this Section, we assess the relative amount of scatter in the
various mass-observable relations derived above. As discussed in
Sec.~\ref{sec:interp}, an observable threshold that yields a higher
best-fitting mass has a mass relation with lower scatter. We define
thresholds corresponding to cluster comoving number densities of
$\bar{n} = \{20, 10, 5, 2.5\} \times 10^{-7} \mpch$. This translates to taking the top \{384, 192, 96, 48\} clusters for the \zbina {} sample, and the top \{456, 233, 116, 58\} clusters for the \zbinb {} sample. We measure the stacked weak lensing signal for each threshold in \ngal, \ltot, and \lbcg {} and compare the derived best-fitting masses in Fig.~\ref{fig:mass_scatter}.

Out of the three parameters considered, we find that \lbcg {} is the poorest tracer of cluster mass. 
This statement is robust to the selection effect introduced by the $N_{200} \ge 10$ cut, since 
the inclusion of poorer, low-mass clusters into the threshold would further decrease the lensing signal. 
We note that the scatter in the mass relation is a combination of intrinsic
and observational scatter, and the contribution from the latter may be significant because
of the difficulty in measuring accurate BCG luminosities. For example, systematic errors 
from sky subtraction are important for BCGs because they have 
large, diffuse envelopes, and deblending issues are also important
because BCGs are located in dense environments.

Figure~\ref{fig:mass_scatter} shows that the best-fitting masses \mnfw {} for clusters in \ngal {} and \ltot {} thresholds 
at the same number density tend to be comparable. However, about 70--80 per cent of the clusters selected by the 
\ngal {} threshold is also selected by the corresponding \ltot {} threshold. Thus, the error bars in these 
data points are tightly correlated, and the differences in the masses are more significant than what one
would estimate by eye. We therefore assess the statistical significance of these differences
using results from many bootstrap datasets. We find that the \ngal {} threshold 
yields a higher mass than the \ltot {} threshold in \{72, 45, 93, 68\} per cent of the cases (for the \zbina {} sample), 
and for \{37, 68, 27, and 90\} per cent of the cases (for the \zbinb {} sample), 
in order of decreasing number density. These high values indicate that \ngal {}
picks out more of the most massive clusters most of the time, and
therefore has smaller scatter than \ltot {} at this range of masses. 
Figure~\ref{fig:mass_scatter} also shows for comparison the masses obtained from thresholds in 
the combined mass tracers, which we discuss in Sec.~\ref{subsec:results_combined_scatter}.

\subsection{Combined mass tracers}
\label{subsec:results_combined}

\begin{figure*} 
\begin{minipage}{170mm}
\includegraphics[width=84mm]{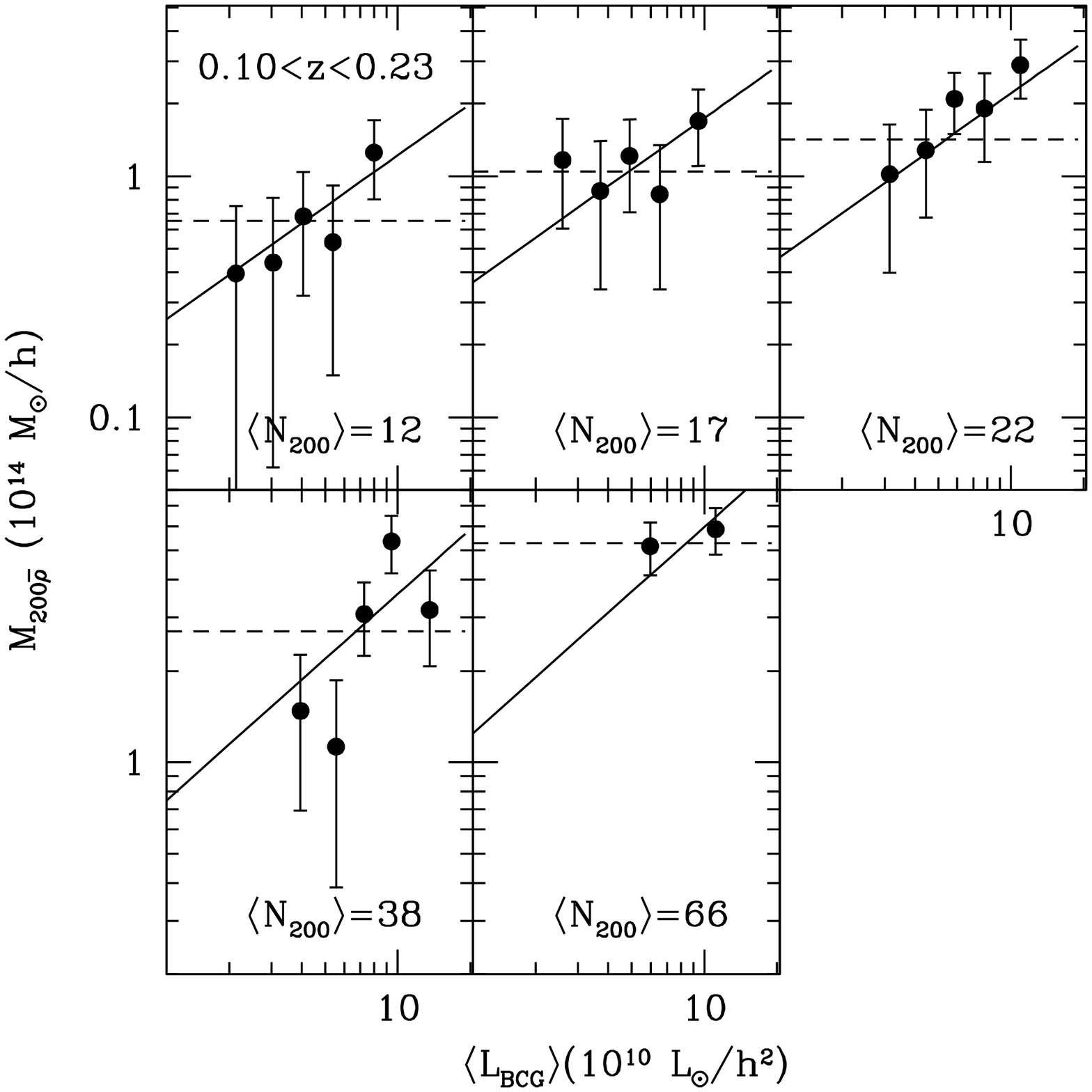}
\includegraphics[width=84mm]{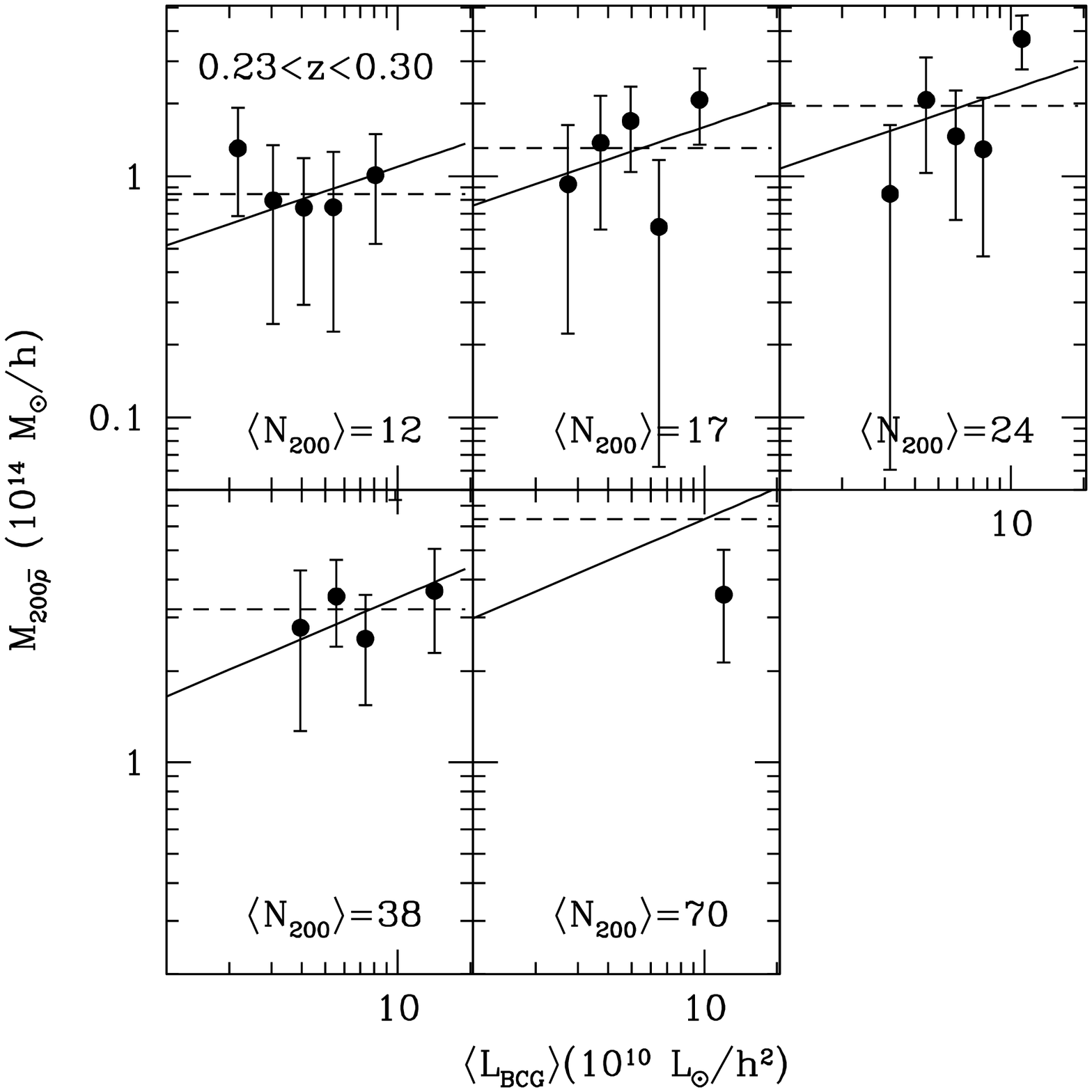}
\caption{Scaling of mean cluster mass \mnfw {} with \lbcg {} within narrow bins in \ngal. The best-fitting mass relation $M(N_{200},L_{\rm BCG})$ (given by Eq.~\ref{eq:2d1}) are shown in solid lines. The mass relation without the \lbcg {} dependence (i.e., with $\gamma_N = 0$) are shown in dashed lines. We find residual scaling with $\gamma_N = 0.71 \pm 0.14$ in the lower redshift sample (left), and with $\gamma_N = 0.34 \pm 0.24$ in the higher redshift sample (right).} 
\label{fig:mass_nlbins}
\end{minipage}
\end{figure*}

In this section, we consider whether adding information from BCG luminosity can provide improved estimates of cluster masses. Our previous analysis shows that \lbcg {} by itself does not trace mass as well as \ngal {} or \ltot. However, the scatter in \lbcg {} at a fixed \ngal {} or \ltot {} suggests that there may be residual scaling of mass with \lbcg. Figure~\ref{fig:mass_bestfit_combined} shows that the scaling of mass with a combination of \ngal {} (or \ltot) and \lbcg {} (lower panels) is tighter than that with \ngal {} or \ltot {} taken alone (upper panels), regardless of the parameter used for binning the clusters. This suggests that the additional information in \lbcg {} reduces the scatter in the mass relation. Here, we consider power law combinations of \lbcg {} with \ngal {} (or \ltot) as mass tracers. We calibrate the mass relation in Sec.~\ref{subsec:results_combined_calib} and assess the scatter in this relation in Sec.~\ref{subsec:results_combined_scatter}.

\subsubsection{Calibration of mean mass-observable relations}
\label{subsec:results_combined_calib}

To consider the scaling of mass with both \ngal {} and \lbcg {} simultaneously, we divide the cluster sample into five bins in \ngal {} and further split these bins in \lbcg, for a total of 22 bins in the two-dimensional \ngal--\lbcg {} space. We make a similar division in \ltot--\lbcg {} space for clusters with $L_{200} > 28$ (for which the sample is complete) resulting in nine bins. We then measure the stacked weak lensing signal around clusters in each bin. We do this analysis for two redshift ranges, \zbina {} and \zbinb. 

We parametrize the scaling of mass as a power law in \ngal (or \ltot) with an additional scaling with \lbcg {} at fixed \ngal {} (or \ltot):
\begin{subeqnarray} \label{eq:mass_2d}
\slabel{eq:2d1}
M_{14}(N_{200},L_{\rm BCG})=M_N^0 (N_{200}/20)^{\alpha_N}(L_{\rm BCG}/\bar{L}_{\rm BCG}^{(N)})^{\gamma_N}\\
\slabel{eq:2d2}
M_{14}(L_{200},L_{\rm BCG})=M_L^0 (L_{200,10}/40)^{\alpha_L}(L_{\rm BCG}/\bar{L}_{\rm BCG}^{(L)})^{\gamma_L}
\end{subeqnarray}
where $M_{14}$ is \mnfw {} in units of $10^{14} h^{-1} M_\odot$, $L_{200,10}$ is the cluster luminosity in units of $10^{10} h^{-2} L_\odot$, and the BCG luminosity dependence is pivoted at the mean \lbcg {} at the given \ngal {} (or \ltot). Parametrizing this mean relation as a power law, the best-fitting relations are: 
\begin{subeqnarray} \label{eq:meanbcg}
\bar{L}_{\rm BCG}^{(N)} \equiv \bar{L}_{\rm BCG}(N_{200}) = a_N N_{200}^{b_N} \\
\bar{L}_{\rm BCG}^{(L)} \equiv \bar{L}_{\rm BCG}(L_{200}) = a_L L_{200,10}^{b_L}         
\end{subeqnarray}
where $a_N = (1.54,1.64) \times 10^{10} h^{-2} L_\odot$, $b_N = (0.41,0.43)$ and $a_L = (0.61,0.58) \times 10^{10} h^{-2} L_\odot$, $b_L = (0.67,0.66)$ for the two redshift ranges ($0.10<z<0.23$, $0.23<z<0.30$). Combining Eqs.~\ref{eq:mass_2d} and \ref{eq:meanbcg} gives a cluster mass estimate for any cluster with measured \ngal {} (or \ltot) and \lbcg.

We derive best-fitting parameters $M^0$, $\alpha$ and $\gamma$ (shown in Table~\ref{tab:chi2fit}) by minimizing $\chi^2$ simultaneously for the set of measured lensing signals. To obtain confidence intervals on these fits, we repeat the fitting procedure for the 1000 bootstrap-resampled datasets, using the analytical covariance matrix (rather than the full bootstrap covariance matrix, which is too noisy to use to weight the fits). The bootstrap-resampled datasets yield Gaussian probability distributions in $M^0$, $\alpha$ and $\gamma$; the 1$\sigma$ errors, and correlation coefficients for these parameters are also shown in Table~\ref{tab:chi2fit}.

Comparison of the best-fitting mass relations for the two redshift
ranges suggests an increase in cluster mass with redshift at fixed
richness. Using the 1000 bootstrap resampled datasets, we find that
the mass normalization for the higher redshift sample is larger than
that for the lower redshift sample at $\sim$97 per cent CL. We note
however that the redshift dependence may result from systematic
effects due to photo-$z$ errors, which have a larger dispersion at
lower redshifts, and/or from evolution in the richness estimator
$N_{200}$ (e.g., due to an incorrect assumption of the evolution of
the luminosity cut $0.4 L_*$). Disentangling these effects from
``true'' evolution requires a more careful control of the
systematics. Hints of an increase in cluster mass with redshift at
fixed \ngal {} have been found in measurements of X-ray luminosities
\citep{2007arXiv0709.1158R} and velocity dispersions
\citep{2007ApJ...669..905B} of clusters in the maxBCG catalog, but no
evidence of evolution had been detected in a previous analysis of
their weak lensing signal \citep{2007arXiv0709.1153S}. 

\begin{table*}
\centering
\begin{tabular}{ccccccc}
\hline
& $M_N^0$ & $\alpha_N$ & $\gamma_N$ & $r(M_N^0,\alpha_N)$ & $r(M_N^0,\gamma_N)$ & $r(\alpha_N,\gamma_N)$ \\
\hline
$0.10<z<0.23$ & $1.27 \pm 0.08$ & 1.20 $\pm$ 0.09 & 0.71 $\pm$ 0.14 & $-0.24$ & $-0.40$ & $0.03$ \\
$0.23<z<0.30$ & $1.57 \pm 0.14$ & 1.12 $\pm$ 0.15 & 0.34 $\pm$ 0.24 & $-0.07$ & $-0.18$ & $0.09$ \\
\hline
& $M_L^0$ & $\alpha_L$ & $\gamma_L$ & $r(M_L^0,\alpha_L)$ & $r(M_L^0,\gamma_L)$ & $r(\alpha_L,\gamma_L)$ \\
\hline
$0.10<z<0.23$ & $1.81 \pm 0.15$ & 1.27 $\pm$ 0.17 & 0.40 $\pm$ 0.23 & $-0.34$ & $-0.17$ & $0.34$ \\
$0.23<z<0.30$ & $1.76 \pm 0.22$ & 1.30 $\pm$ 0.29 & 0.26 $\pm$ 0.41 & $-0.42$ & $-0.35$ & $0.41$ \\
\hline
\end{tabular}
\caption{Best-fitting parameters for the scaling of cluster mass with \ngal {} and \lbcg {} (Eq.~\ref{eq:2d1}), and with \ltot {} and \lbcg {} (Eq.~\ref{eq:2d2}). 
The 1$\sigma$ errors and correlation coefficients $r$ in the table are derived from 1000 bootstrap-resampled datasets.} 
\label{tab:chi2fit} \end{table*}

\begin{figure} 

\includegraphics[width=84mm]{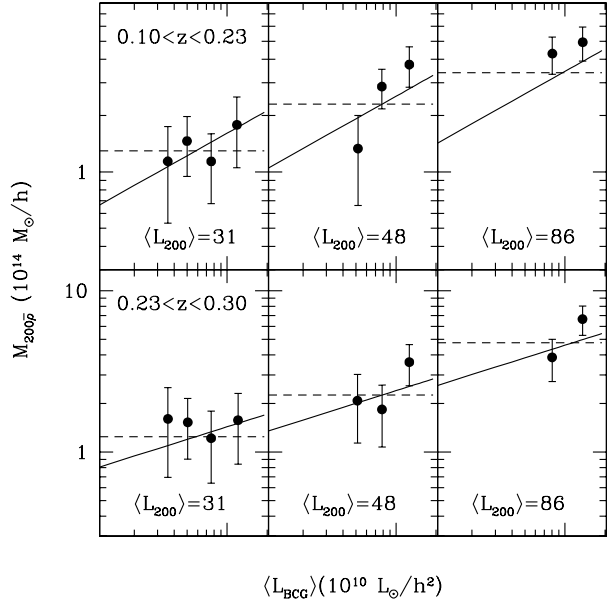}
\caption{Scaling of mean cluster mass \mnfw {} with \lbcg {} within
narrow bins in \ltot. The mean \ltot {} in each bin is shown in units
of $10^{10} h^{-2} L_\odot$; we restrict this analysis to $L_{200} >28
\times 10^{10} h^{-2} L_\odot$, for which the sample is complete. We
find residual scaling of \mnfw {} with \lbcg {} at fixed \ltot {} in
the lower redshift sample (upper panels), with $\gamma_L = 0.40 \pm
0.23$ ($\sim 2\sigma$) and no significant evidence for residual scaling in 
the higher redshift sample (lower panels), with $\gamma_L = 0.26 \pm 0.41$.}
\label{fig:mass_llbins}

\end{figure}

Figures~\ref{fig:mass_nlbins} and \ref{fig:mass_llbins} show the scaling of cluster mass \mnfw {} with \lbcg {} within narrow bins in \ngal {} and \ltot. These scalings are traced well by the best-fitting relations Eqs.~\ref{eq:2d1} and \ref{eq:2d2}. 
At fixed \ngal, residual scaling with \lbcg {} is seen with $\gamma_N = 0.71 \pm 0.14$ ($\sim 5\sigma$) for
the lower redshift sample, and with $\gamma_N = 0.34 \pm 0.24$ for the higher redshift sample.
At fixed \ltot, we find $\gamma_L = 0.40 \pm 0.23$ ($\sim 2\sigma$) for the lower redshift sample, and
$\gamma_L = 0.26 \pm 0.41$ for the higher redshift sample. Constraints for the scaling with \ltot {} are
relatively weaker because of the luminosity cut applied to the complete sample, which reduces the number 
of clusters to about one-third of the full sample. The scaling parameters are less well-constrained
for the higher redshift range because there are fewer lensed sources behind the high-redshift clusters. 

\subsubsection{Scatter in the mass-observable relations}
\label{subsec:results_combined_scatter}

We turn to the question of whether exploiting information about BCG luminosity in addition to either \ngal {} or \ltot {} reduces the scatter in the mass relation. Similar to Sec.~\ref{subsec:results_indiv_scatter}, we rank clusters according to \ltot\lbcg$^{\beta_L}$ {} and \ngal\lbcg$^{\beta_N}$ and take the top $N$ clusters to define thresholds with comoving number densities $\bar{n} = \{20, 10, 5, 2.5\} \times 10^{-7} \mpch$. We explore a set of values of exponents, $\beta_N = \{0.25, 0.50, 0.75, 1.0, 1.5, 2.0\}$ and $\beta_L = \{0.2, 0.4, 0.6, 0.8, 1.0\}$, to find the one that maximizes \mnfw, or equivalently, minimizes the scatter in the mass-observable relation. We do this analysis for two redshift ranges, \zbina {} and \zbinb.

The exponents that yield the highest masses at each number density are (from highest to lowest number density): $\beta_N^{\rm (best)} = \{1.5, 1.5, 0.25, 0.25\}$ and $\beta_L^{\rm (best)} = 0.4$ for the lower redshift sample and $\beta_N^{\rm (best)}=\{1.5, 1.0, 0.5, 0.5\}$ and $\beta_L^{\rm (best)}= \{0.8, 0.6, 0.6, 0.2\}$ for the higher redshift sample. In general, the tracer with the minimal scatter is a combination of \ngal {} and \lbcg {} (except for $\bar{n}=2.5 \times 10^{-7} \mpch$ in the higher redshift sample, where \ngal {} alone yields the highest mass; one possible reason for this trend is that at higher redshifts, the large \lbcg {} bins are more likely to be contaminated by low luminosity objects for which the photo-$z$ has been overestimated (Sec.~\ref{subsec:sys_photoz})).

The error bars are tightly correlated between the combined and individual tracers, as well as between different $\beta_N$ or $\beta_L$ values, because a significant fraction of the clusters that satisfy the different thresholds are the same. For example, for the lowest number density bin $\bar{n} = 2.5 \times 10^{-7} \mpch$, there is substantial overlap between clusters satisfying the threshold in $\beta_N^{\rm (best)}$ and in \ngal {} (94 per cent), \ltot {} (83 per cent), and \lbcg {} (26 per cent). We assess the significance of the differences in the masses using the 1000 bootstrap-resampled datasets. We find that the combined tracers with exponents $\beta_N^{\rm (best)}$ yield higher masses than \ngal, \ltot, or \lbcg {} in the majority of cases ($>50$ per cent), for the range of number densities we consider.

We emphasize that this result is 
relevant even if we are not complete in \lbcg {} or \ltot, in the sense that this is 
the estimate that minimizes the scatter among the clusters we have.
This does not imply that we could not have an even better sample if we 
included clusters with \ngal $<10$ for which \lbcg {} is high. 
However, from Fig.~\ref{fig:correl}, we see that we are 
complete for $N_{200}L_{{\rm BCG},10}^{0.75}>80$, so our results are not affected by incompleteness 
for number densities below $5\times 10^{-6} \mpch$. 

Together with the results of Sec.~\ref{subsec:results_combined_calib}, these findings suggest that additional information from \lbcg {} provides improved determination of cluster masses, both in the mean and the scatter of the mass-observable relation. 

\section{Summary and Conclusions}
\label{sec:conclusions}

We considered optical parameters that are available in large samples of clusters of galaxies: cluster richness \ngal, cluster luminosity \ltot, and the luminosity of the brightest cluster galaxy \lbcg, as well as power law combinations of \ngal {} with \lbcg, and \ltot {} with \lbcg, to determine which is the best mass tracer for clusters. 

We calibrate the mean mass relation for these tracers by measuring the
stacked weak lensing signal around clusters rank ordered according to a given parameter. 
Our best-fitting mass relations for \ngal {} and \ltot {} are given in Eqs.~\ref{eq:scal1} and \ref{eq:scal2}. 
We then ask whether the weak lensing signal changes significantly 
when a second parameter is added to the first one. 
We can exploit any such residual scaling to derive improved, lower-scatter mass 
tracers. We explore such tracers in 
the form \ngal$^{\alpha_N}$\lbcg$^{\gamma_N}$ and \ltot$^{\alpha_L}$\lbcg$^{\gamma_L}$. 
The best-fitting mass relations are given in Eqs.~\ref{eq:2d1} and \ref{eq:2d2}, with parameters given in Table~\ref{tab:chi2fit}.
The best mass tracer \mnfw {} (in units of $10^{14} h^{-1} M_\odot$) we find is (for the lower redshift sample):
\[
M_{14} = (1.27 \pm 0.08) \left(\frac{N_{200}}{20}\right)^{1.20 \pm 0.09} \left(\frac{L_{\rm BCG}}{\bar{L}_{\rm BCG}(N_{200})}\right)^{0.71 \pm 0.14}
\]
where $\bar{L}_{\rm BCG}(N_{200}) = 1.54~N_{200}^{0.41} \times 10^{10}
h^{-2} L_\odot$ is the mean BCG luminosity at a given richness. 

Our results suggest that \lbcg {} is an important second parameter in addition to \ngal {} and \ltot.
At fixed \ngal, residual scaling with \lbcg {} is seen at the $\sim 5\sigma$ level 
in the lower redshift sample (\zbina), and at the $\sim 1.5\sigma$ level in the higher redshift sample (\zbinb).
The need for a second parameter is less evident when \ltot {} is used as the primary variable instead of \ngal;
we find that residual scaling with \lbcg {} is preferred at the $\sim 2\sigma$ level in the lower redshift sample, and
find no evidence for residual scaling in the higher redshift sample.

We assess the relative amount of scatter in 
the various mass-observable relations by measuring the stacked weak lensing signal around clusters satisfying thresholds in each parameter. 
For a given comoving number density of clusters, low-scatter mass tracers will select more of the 
most massive clusters in the sample and thus yield 
a stronger lensing signal, 
compared to a large-scatter mass tracer. Among the parameters \ngal, \ltot, and \lbcg, cluster richness is the best mass tracer for clusters, while \lbcg {} is the poorest tracer. We find that a combined tracer of the form \ngal\lbcg$^{\beta_N}$ reduces the scatter in the mass relation compared to cluster richness taken alone, for the most massive clusters in the sample.

From SDSS spectroscopy of clusters in the maxBCG catalog, \citet{2007ApJ...669..905B} found residual
scaling of velocity dispersions with BCG luminosity \lbcg {} at fixed richness
\ngal. Our results consequently confirm that this residual scaling
also appears in the projected mass distributions.  

Our results are consistent with the current picture of
cluster formation from halo mergers. $N$-body simulations and semi-analytic models find
that at a fixed mass, dark matter haloes which form earlier have
brighter, redder central sub-haloes (i.e., brighter, redder BCGs) and
lower richness \citep{2006ApJ...652...71W,2007MNRAS.374.1303C}. This
may result from the satellites having had more time to merge onto the
BCG, lowering the richness from when the cluster formed while
enhancing the BCG luminosity.  This implies that \ngal {} and \lbcg {} are anti-correlated at fixed mass, and provides an explanation for our result above, i.e., that a combination of these two observables yields a
tighter relation with mass than either of them taken alone. 

The weaker residual scaling with \lbcg {} when using \ltot {} instead of \ngal, 
suggests that the anti-correlation between \ltot {} and
\lbcg {} at fixed mass is much weaker; this is also consistent with the above
scenario, since the luminosity of the BCG is included in the cluster luminosity. 
Moreover, this result constrains the amount of light that has been lost to the intra-cluster medium due to 
the merging of red satellite galaxies with the BCG since the formation of the cluster. 
If this was a significant fraction of the cluster luminosity in red galaxies, \ltot {} 
would be lower for earlier-forming clusters, and therefore anti-correlated with \lbcg.
We do not detect such an effect, so our results are consistent with a scenario where
the cluster luminosity in red galaxies remains approximately constant over
time.

Independent of the underlying astrophysical mechanisms, the improved
mass tracers we found can be used to obtain accurate mass estimates and
define mass thresholds in cluster samples with optical
data. These in turn can be used to provide more precise constraints on
cosmological parameters, such as the amplitude of mass fluctuations
$\sigma_8$, which will be the subject of  future work.

\section*{Acknowledgements}

We thank Ben Koester, Tim McKay, Erin Sheldon,
and Risa Wechsler for useful
discussions regarding the maxBCG cluster catalog.
R.M. is supported by NASA 
through Hubble Fellowship grant \# HST-HF-01199.02-A awarded by the
Space Telescope Science Institute, which is operated by the
Association of Universities for Research in Astronomy, Inc., for NASA, 
under contract NAS 5-26555. C.H. is supported by the U.S. Department of Energy under contract DE-FG03-02-ER40701. U.S. is supported by the
Packard Foundation 
and NSF CAREER-0132953.

\bibliography{cosmo,cosmo_preprints}
\end{document}